\pdfoutput=1
\PassOptionsToPackage{table}{xcolor}
\documentclass[sigconf,screen]{acmart}

\usepackage{multirow}
\usepackage{dblfloatfix}
\usepackage{enumitem}
\usepackage{bigstrut}
\usepackage[tikz]{bclogo}
\usepackage[skins]{tcolorbox}
\usepackage{wrapfig}
\usepackage[flushleft]{threeparttable}
\usepackage{float}
\usepackage{adjustbox}
 
\definecolor{deeppink}{HTML}{D28986} 
\definecolor{pink}{HTML}{FFABA7} 
\definecolor{lightpink}{HTML}{FFCCC9} 
\usepackage[usestackEOL]{stackengine}
\usepackage{xcolor}

\usepackage[final]{listings}
\usepackage[linesnumbered,ruled,vlined]{algorithm2e}

\definecolor{deeppink}{HTML}{D28986} 
\definecolor{pink}{HTML}{FFABA7} 
\definecolor{lightpink}{HTML}{FFCCC9}
\definecolor{lightgreen}{HTML}{9AFF99}
\definecolor{green}{HTML}{34FF34}
\usepackage[usestackEOL]{stackengine}
\usepackage{xcolor}

\SetCommentSty{mycommfont}

\SetKwInput{KwInput}{Input}                % Set the Input
\SetKwInput{KwOutput}{Output}              % set the Output

\SetKwFunction{FMain}{Main}
\SetKwFunction{FSum}{Sum}
\SetKwFunction{FSub}{Fair-SMOTE}

\AtBeginDocument{%
  \providecommand\BibTeX{{%
    \normalfont B\kern-0.5em{\scshape i\kern-0.25em b}\kern-0.8em\TeX}}}

\newcommand{\IT}[1]{{\bf%
DODGE(\ifx*#1$\mathcal{E}$\else#1\fi)}}

\newcommand{\bi}{\begin{itemize}[leftmargin=0.4cm]}
	\newcommand{\ei}{\end{itemize}}
\newcommand{\be}{\begin{enumerate}[leftmargin=0.4cm]}
	\newcommand{\ee}{\end{enumerate}}

\begin{document}
% \title{Fair-SSL: Achieving Fairness using Semi-Supervised Learning}
\title{Fair-SSL: Building fair ML Software with less data}

% Zhi-Wu Zhang, Xiao-Yuan Jing, and Tie-Jian Wang. Label propagation basedsemi-supervised learning for software defect prediction.Automated SoftwareEngineering, 2017.

\author{Joymallya Chakraborty}
\email{jchakra@ncsu.edu}
\affiliation{%
\institution{North Carolina State University}
\city{Raleigh}
\country{USA}}

\author{Suvodeep Majumder}
\email{smajumd3@ncsu.edu}
\affiliation{%
\institution{North Carolina State University}
\city{Raleigh}
\country{USA}}

\author{Huy Tu}
\email{hqtu@ncsu.edu}
\affiliation{%
\institution{North Carolina State University}
\city{Raleigh}
\country{USA}}

\begin{abstract}
Ethical bias in machine learning models has become a matter of concern in the software engineering community. Most of the prior software engineering works concentrated on finding ethical bias in models rather than fixing it. After finding bias, the next step is mitigation. Prior researchers mainly tried to use supervised approaches to achieve fairness. However, in the real world, getting data with trustworthy ground truth is challenging and also ground truth can contain human bias. 

Semi-supervised learning is a technique where, incrementally, labeled data
is used to generate pseudo-labels for the rest of data
(and then all that data is used for
  model training).
In this work, we apply four popular semi-supervised techniques as pseudo-labelers to create fair classification models. Our framework,  \textbf{Fair-SSL},  takes a very small amount (10\%) of labeled data as input and generates pseudo-labels for the unlabeled data. We then synthetically generate new data points to balance the training data based on class and protected attribute as proposed by Chakraborty et al. in FSE 2021. Finally, classification model is trained on the balanced pseudo-labeled data and validated on test data. After experimenting on ten datasets and three learners, we find  that Fair-SSL achieves similar performance as three  
state-of-the-art bias mitigation algorithms. That said, the clear advantage of Fair-SSL is that   it requires only 10\% of the labeled training data. 

To the best of our knowledge,  this is the first SE work where semi-supervised techniques are used to fight against ethical bias in SE ML models. To facilitate open science and replication, all our  source code and datasets are publicly available  at \url{https://github.com/joymallyac/FairSSL}.

\end{abstract}


\copyrightyear{2022}
\acmYear{2022}
\setcopyright{acmcopyright}\acmConference[FairWare '22 ]{International Workshop on Equitable Data and Technology}{May 9, 2022}{Pittsburgh, PA, USA}
\acmBooktitle{International Workshop on Equitable Data and Technology (FairWare '22 ), May 9, 2022, Pittsburgh, PA, USA}
\acmPrice{15.00}
\acmDOI{10.1145/3524491.3527305}
\acmISBN{978-1-4503-9292-1/22/05}

\maketitle

\section{Introduction}

Machine learning software has become ubiquitous in our society. Software is making autonomous decisions in criminal sentencing~\cite{10.1145/3322640.3326705}, loan approvals~\cite{forbes}, patient diagnosis~\cite{Medical_Diagnosis}, hiring candidates~\cite{hireview}, and whatnot. It is the duty of software researchers and engineers to produce high-quality software that always makes fair decisions. However, in recent times, there are numerous examples where machine learning software is found to have biased behavior based on some protected attributes like sex, race, age, marital status, etc. Google translate, the most popular translation engine in the world, shows gender bias. ``She is an engineer, He is a nurse'' is translated into Turkish and then again into English becomes ``He is an engineer, She is a nurse''~\cite{Caliskan183}. YouTube makes more mistakes when it automatically generates closed captions for videos with female than male voices~\cite{tatman-2017-gender}. Amazon's automated recruiting tool was found to be biased against women~\cite{Amazon_Recruit}.  A widely used face recognition software was found to be biased against dark-skinned women~\cite{Skin_Bias}. Angel et al. commented that software showing bias is considered as poor quality software and should not be used in real life applications~\cite{Angell:2018:TAT:3236024.3264590}. It is time for software engineering researchers to dive into the field of software fairness and try to build fairer software to prevent discriminative behaviors. 

\begin{table*}[!b]
\caption{Details of the datasets used in this research.}
\label{datasets}
% \small
\footnotesize
\begin{tabular}{|c|c|c|cc|cc|}
\hline
\rowcolor[HTML]{C0C0C0} 
\textbf{Dataset}    & \textbf{\#Rows} & \textbf{\#Cols} & \multicolumn{2}{c|}{\cellcolor[HTML]{C0C0C0}\textbf{Protected Attribute}}                         & \multicolumn{2}{c|}{\cellcolor[HTML]{C0C0C0}\textbf{Class Label}}                      \\ \hline
\rowcolor[HTML]{C0C0C0} 
\textbf{}           & \textbf{}       & \textbf{}       & \multicolumn{1}{c|}{\cellcolor[HTML]{C0C0C0}\textbf{Privileged}} & \textbf{Unprivileged}          & \multicolumn{1}{c|}{\cellcolor[HTML]{C0C0C0}\textbf{Favorable}} & \textbf{Unfavorable} \\ \hline
Adult Census~\cite{ADULT}        & 48,842          & 14              & \multicolumn{1}{c|}{Sex-Male; Race-White}                        & Sex-Female; Race-Non-white     & \multicolumn{1}{c|}{High Income}                                & Low Income           \\ \hline
Compas~\cite{COMPAS}             & 7,214           & 28              & \multicolumn{1}{c|}{Sex-Male; Race-Caucasian}                    & Sex-Female; Race-Not Caucasian & \multicolumn{1}{c|}{Did not reoffend}                           & Reoffended           \\ \hline
German Credit~\cite{GERMAN}      & 1,000           & 20              & \multicolumn{1}{c|}{Sex-Male}                                    & Sex-Female                     & \multicolumn{1}{c|}{Good Credit}                                & Bad Credit           \\ \hline
Default Credit~\cite{DEFAULT}      & 30,000          & 23              & \multicolumn{1}{c|}{Sex-Male}                                    & Sex-Female                     & \multicolumn{1}{c|}{Default Payment-Yes}                        & Default Payment-No   \\ \hline
Heart Health~\cite{HEART}        & 297             & 14              & \multicolumn{1}{c|}{Age-Young}                                   & Age-Old                        & \multicolumn{1}{c|}{Not Disease}                                & Disease              \\ \hline
Bank Marketing~\cite{BANK}      & 45,211          & 16              & \multicolumn{1}{c|}{Age-Old}                                     & Age-Young                      & \multicolumn{1}{c|}{Term Deposit - Yes}                         & Term Deposit - No    \\ \hline
Home Credit~\cite{HOME_CREDIT}         & 3,075,11        & 240             & \multicolumn{1}{c|}{Sex-Male}                                    & Sex-Female                     & \multicolumn{1}{c|}{Approved}                                   & Rejected             \\ \hline
Student Performance~\cite{STUDENT} & 1,044           & 33              & \multicolumn{1}{c|}{Sex-Male}                                    & Sex-Female                     & \multicolumn{1}{c|}{Good Grade}                                 & Bad Grade            \\ \hline
MEPS15~\cite{MEPS15}              & 35,428          & 1,831           & \multicolumn{1}{c|}{Race-White}                                  & Race-Non-white                 & \multicolumn{1}{c|}{Good Utilization}                           & Bad Utilization      \\ \hline
MEPS 16~\cite{MEPS16}             & 34,656          & 1,941           & \multicolumn{1}{c|}{Race-White}                                  & Race-Non-white                 & \multicolumn{1}{c|}{Good Utilization}                           & Bad Utilization      \\ \hline
\end{tabular}
\end{table*}

\begin{table*}[]
% \small
\footnotesize
\caption{Definition of the performance and fairness metrics used in this study.}
\label{metrics_table}
\begin{tabular}{|l|c|l|c|}
\hline
\rowcolor[HTML]{C0C0C0} 
\multicolumn{1}{|c|}{\cellcolor[HTML]{C0C0C0}{\color[HTML]{333333} \textbf{Performance Metric}}}                                           & {\color[HTML]{333333} \textbf{\begin{tabular}[c]{@{}c@{}}Ideal \\ Value\end{tabular}}} & \multicolumn{1}{c|}{\cellcolor[HTML]{C0C0C0}{\color[HTML]{333333} \textbf{Fairness Metric}}}                                                                                                                                                                                                                                                                            & {\color[HTML]{333333} \textbf{\begin{tabular}[c]{@{}c@{}}Ideal \\ Value\end{tabular}}} \\ \hline
Recall = TP/P = TP/(TP+FN)                                                                                                                 & 1                                                                                      & \begin{tabular}[c]{@{}l@{}}Average Odds Difference (AOD): Average of difference in False Positive Rates(FPR) and True Positive Rates(TPR) \\ for unprivileged and privileged groups\\ TPR = TP/(TP + FN),  FPR = FP/(FP + TN), AOD = {[}(FPR\_U - FPR\_P) + (TPR\_U - TPR\_P){]} * 0.5\end{tabular}                                                                     & 0                                                                                      \\ \hline
False alarm = FP/N = FP/(FP+TN)                                                                                                            & 0                                                                                      & \begin{tabular}[c]{@{}l@{}}Equal Opportunity Difference (EOD):  Difference of True Positive Rates(TPR) for unprivileged and privileged \\ groups. EOD = TPR\_U - TPR\_P\end{tabular}                                                                                                                                                                                    & 0                                                                                      \\ \hline
\begin{tabular}[c]{@{}l@{}}Accuracy = $ \frac { (TP + TN) } {(TP + FP + TN + FN)}$\end{tabular} & 1                                                                                      & \begin{tabular}[c]{@{}l@{}}Statistical Parity Difference (SPD): Difference between probability of unprivileged group (PA = 0) gets favorable \\ prediction ($\hat{Y} = 1$) \& probability of privileged group (PA = 1) gets favorable prediction ($\hat{Y} = 1$). \\ $SPD = P[\hat{Y}=1|PA=0] - P[\hat{Y}=1|PA=1]$\end{tabular} & 0                                                                                      \\ \hline
Precision = TP/(TP+FP)                                                                                                                     & 1                                                                                      & \begin{tabular}[c]{@{}l@{}}Disparate Impact (DI): Similar to SPD but instead of the difference of probabilities, the ratio is measured. \\ $DI = P[\hat{Y}=1|PA=0]/P[\hat{Y}=1|PA=1]$\end{tabular}                                                                                                                              & 1                                                                                      \\ \hline
F1 Score = $ \frac {2 * (Precision * Recall) }{(Precision + Recall)}$                                                                      & 1                                                                                      &                                                                                                                                                                                                                                                                                                                                                                         &                                                                                        \\ \hline
\end{tabular}
\end{table*}

% \vspace{-1cm}

A machine learning software can acquire bias in various ways~\cite{Brun2018SoftwareF}. Prior studies ~\cite{jiang2019identifying,chen2018classifier} mentioned that most of the time bias comes from the training data. If training data contains improper labels, that bias gets induced into model while training. In an ACM SIGSOFT Distinguished award winning paper, Chakraborty et al.~\cite{Chakraborty2021BiasIM} found out that bias comes from improper data labels and imbalanced data distribution. They said if the training data contains more examples of a certain group getting privileged (males being hired for a job) and another group getting betrayed (females getting more rejections); the machine learning model acquires that bias while training and in the future makes unfair predictions. Their algorithm, Fair-SMOTE, improved both the fairness and performance of the model and broke the premise of Berk et al.~\cite{berk2017fairness} who claimed \textit{``It is impossible to achieve fairness and high performance simultaneously (except in trivial cases)''}.

We consider Fair-SMOTE~\cite{Chakraborty2021BiasIM} as our baseline method. It is a supervised approach and uses 100\% training data labels. But gathering good quality labeled data is very challenging. Human labeling is an extremely costly process ~\cite{3dsignal,googlecloud, tu2020better, tu2021frugal} and there is a high possibility of human bias getting injected into the training data~\cite{JamesManyika,MarkXiang}. That said, blindly trusting ground truth labels may induce bias in the machine learning model. Hence, it is timely to ask:
\begin{quote}
    {\em Can we reduce the labeling effort associated with building fair models?}
\end{quote}
In this work, we try to answer that question by using semi-supervised learning~\cite{Zhu06semi-supervisedlearning} that works with a small amount of labeled data and a large amount of unlabeled data. We build a framework called \textbf{Fair-Semi-Supervised-Learning (Fair-SSL)} that uses four state-of-the-art semi-supervised techniques - self-training, label propagation, label spreading, \& co-training. Fair-SSL is a pseudo-labeling framework. It learns from the combination of labeled \& unlabeled data and then pseudo-labels the unlabeled data. Results show that Fair-SSL performs as good as three other state-of-the-art fairness algorithms~\cite{Chakraborty_2020,Chakraborty2021BiasIM,NIPS2017_6988}. That means even if available ground truth is corrupted or a very few labeled data points are available initially, fairness could still be achievable. Overall, this paper makes the following contributions:

\begin{itemize}
    \item This is the first SE work using semi-supervised learning to generate fair classification models. 
    \item Fair-SSL works with a very small amount of labeled training data (10\%). Thus, we can avoid the costly process of data labeling. Hence, it is cost effective.
    \item We have shown a technique based on ``situation testing''~\cite{USA_SituationTesting} to create fairly labeled data without using any human intervention.
    \item We have given a comparative analysis of four popular semi-supervised algorithms in the context of software fairness. 
    \item Our results show that semi-supervised algorithms can be used to generate fairer and better performing models.
\end{itemize}

% The rest of the paper is organized as follows. Section \ref{Background} provides an overview of software fairness and prior works. It also describes some fairness related terminology and metrics. Section \ref{methodology} explains the framework of Fair-SSL. Section \ref{results} shows the results for six research questions. Section \ref{discussion} summarizes the results. In section \ref{threats}, we have stated the threats to validity of our work. Finally, Section \ref{conclusion} concludes the paper.
% https://arxiv.org/pdf/2009.12040.pdf \\
% https://arxiv.org/pdf/2009.06190.pdf \\
% https://arxiv.org/pdf/1912.13230.pdf
% https://arxiv.org/pdf/1901.04966.pdf

% Very good video - https://www.youtube.com/watch?v=tVsVmy6w7FE&ab_channel=AnalyticsUniversity

\section{Background}
\label{Background}

\textbf{Software Fairness} - 
Big software industries have started putting more and more importance on ethical issues of ML software. AI Fairness 360~\cite{AIF360} from IBM, Fairlearn~\cite{Fairlearn} from Microsoft are great initiatives to attract bigger audience. The software academic community, in spite of having a delayed start,  is taking initiatives to fight against this critical social bane. ICSE
2018 hosted Fairware~\cite{FAIRWARE}; ASE 2019 organized EXPLAIN~\cite{EXPLAIN}. The IEEE \cite{IEEEethics}, the  European Union \cite{EU} recently published the ethical principles of AI. It is stated that every machine learning software must be fair when it is used in real-life applications. Thus, testing software for bias and mitigating bias have now become an unavoidable step in software life cycle.

Some popular fairness testing tools are THEMIS ~\cite{Angell:2018:TAT:3236024.3264590}, Symbolic Generation~\cite{Aggarwal:2019:BBF:3338906.3338937}, AEQUITAS ~\cite{Udeshi_2018}, and white-box testing tool~\cite{10.1145/3377811.3380331}. In case of bias mitigation, we see very few works in SE venues. As per our knowledge, there are only two frameworks Fairway~\cite{Chakraborty_2020} and Fair-SMOTE~\cite{Chakraborty2021BiasIM} that tried to mitigate bias in SE ML models. Both of them are supervised methods that require a lot of labeled training data. We, in this work, used semi-supervised approach and achieved similar or better performance using only 10\% labeled training data.\\
\textbf{Semi-supervised Learning} - 
Supervised machine learning models, specially the deep learning models, require a huge amount of labeled data for training. Gathering good quality labeled training data is the most expensive part of ML pipeline~\cite{3dsignal,googlecloud}. For labeling purpose, usage of human beings is a very expensive process~\cite{3dsignal,googlecloud}. Even then, human bias may get injected into training data~\cite{JamesManyika,MarkXiang}. 

Semi-supervised learning (SSL) can address all these issues.  
SSL requires a small amount of labeled data to begin with~\cite{Zhu06semi-supervisedlearning}, then using an incremental approach, unlabeled data is pseudo-labeled, and the combined data is used for model training. SSL has been used in various domains of software engineering such as defect prediction~\cite{10.1007/s10515-016-0194-x}, test case prioritization~\cite{10.1145/3338906.3340448}, static warning analysis~\cite{tu2021frugal}, software vulnerability prediction~\cite{8883076}, and many more. To the best of our knowledge, this
paper is the first SE work to try SSL in the context of fairness. Outside of SE,  we found only one study by Zhang et al.~\cite{Zhang_2020_SSL} studying SSL and fairness. Whereas they experimented on only three datasets and used only self-training technique, we evaluated our results based on nine metrics, ten datasets, and three learners.\\
\textbf{Fairness Terminology \& Metrics} - 
At first, we need to define some fairness related terms. Table \ref{datasets} contains ten datasets used in this study. Most of the prior works~\cite{NIPS2017_6988,Galhotra_2017,zhang2018mitigating,Kamiran:2018:ERO:3165328.3165686,chakraborty2019software,9286091} used one or two datasets whereas we used ten of them. All these datasets are binary classification datasets i.e. class labels have two values. 
A class label is called a \textit{favorable label} if it gives an advantage to the receiver such as receiving a loan, being hired for a job.  A \textit{protected attribute} is an attribute that divides the whole population into two groups (privileged \& unprivileged) that have differences in terms of receiving benefits. For example, in \textit{Home Credit} dataset, based on ``sex'', ``male'' is privileged and ``female'' is unprivileged. In the context of classification, the goal of fairness is \textit{giving similar treatment to privileged and unprivileged groups}.

Table \ref{metrics_table} contains the definitions of the five performance metrics and four fairness metrics used in this study. We chose these metrics because they were widely used in the literature~\cite{Chakraborty_2020,Biswas_2020,chakraborty2019software,hardt2016equality,9286091,Chakraborty2021BiasIM}. For recall, precision, accuracy, F1 \& DI {\em larger values} are {\em better}; For false alarm, AOD, EOD, \& SPD {\em smaller values} are {\em better}. 
For readability, while showing results we compute \mbox{abs(1 - DI)} so that all four fairness metrics are lower the better (0 means no bias).
\hspace{-12cm}

\section{Methodology}
\label{methodology}

Fair-SSL contains three major steps: a) Select a small amount of labeled data (10\%)  in a way that initial labeling does not contain bias (see \S\ref{methodology_part_1}); b) Pseudo-label the unlabeled data using semi-supervised approaches (see \S\ref{methodology_part_2}); c) Balance the combined training data (labeled + pseudo labeled) based on protected attribute and class label (see \S\ref{methodology_part_3}). Finally, we train ML models on the generated balanced data and test on the test data. We use 80-20 train-test division and test set is used only for final score reporting. 

\vspace{-0.1cm}

\subsection{Prepare the Fairly Labeled data}
\label{methodology_part_1}
Our first task is preparing the initial fairly labeled set. All the datasets in Table~\ref{datasets} are already labeled. But are those labels fair? Can we just randomly pick one portion of that data as fairly labeled? How much labeled data is required to start with?  We are going to find all the answers soon.

Prior studies~\cite{Chakraborty2021BiasIM,Das2020fairML,jiang2019identifying,Ted_Simons} have experimented with the datasets of Table ~\ref{datasets} and found out that more or less 10\% data labels contain unfair decisions. That means if we randomly pick up some portion of the data, we may end up selecting some improperly labeled rows and training ML models on that corrupted data will introduce bias. Thus, at an early stage, we decided to re-label the data.

In literature, there are mainly two approaches for executing the labeling process. The first one is \textit{manual labeling/crowdsourcing}. The second one is \textit{semi-supervised pseudo-labeling}. At first, we tried to do crowdsourcing for labeling. Our attempt was not successful. The datasets we use in the fairness domain are not very easy to be labeled by common people. Most are financial datasets; Compas is a criminal sentencing dataset; Heart health is a medical dataset. Hence special expertise is needed to label these kinds of data. It was out of our scope to find that kind of experienced people. Also, this manual labeling is a super expensive process. Besides, there is possibility of human bias getting injected into data labels~\cite{JamesManyika,MarkXiang}. Therefore, we discarded the idea of manual labeling.\\
\textbf{Situation Testing}: 
Chakraborty et al. reported that in these datasets, around 10\% of the labels are unfair labels~\cite{Chakraborty2021BiasIM}. Thus we could select only fairly labeled rows from the available data and treat the rest as unlabeled data. To achieve that, we used the concept of \textit{situation testing}~\cite{10.5555/3060832.3061001}. It is a research technique used in the legal field~\cite{USA_SituationTesting} where decision makers' candid responses to applicant’s personal characteristics are captured and analyzed. 

At first, we divide the data based on the protected attribute. Then, we train two different logistic regression models (any other simple statistical model can be used) on those two subgroups (``male'' and ``female''). Then for all the training data points, we check the predictions of these two models. For a particular data point, if the predictions of two models match with the ground truth label, we keep the data point with the same label as it was fairly labeled. If the model predictions contradict with each other or the ground truth, we discard the label and treat the data point as unlabeled.

\begin{figure}[!ht]
\centering
\adjustbox{max height=4in}{
\includegraphics[width=\linewidth]{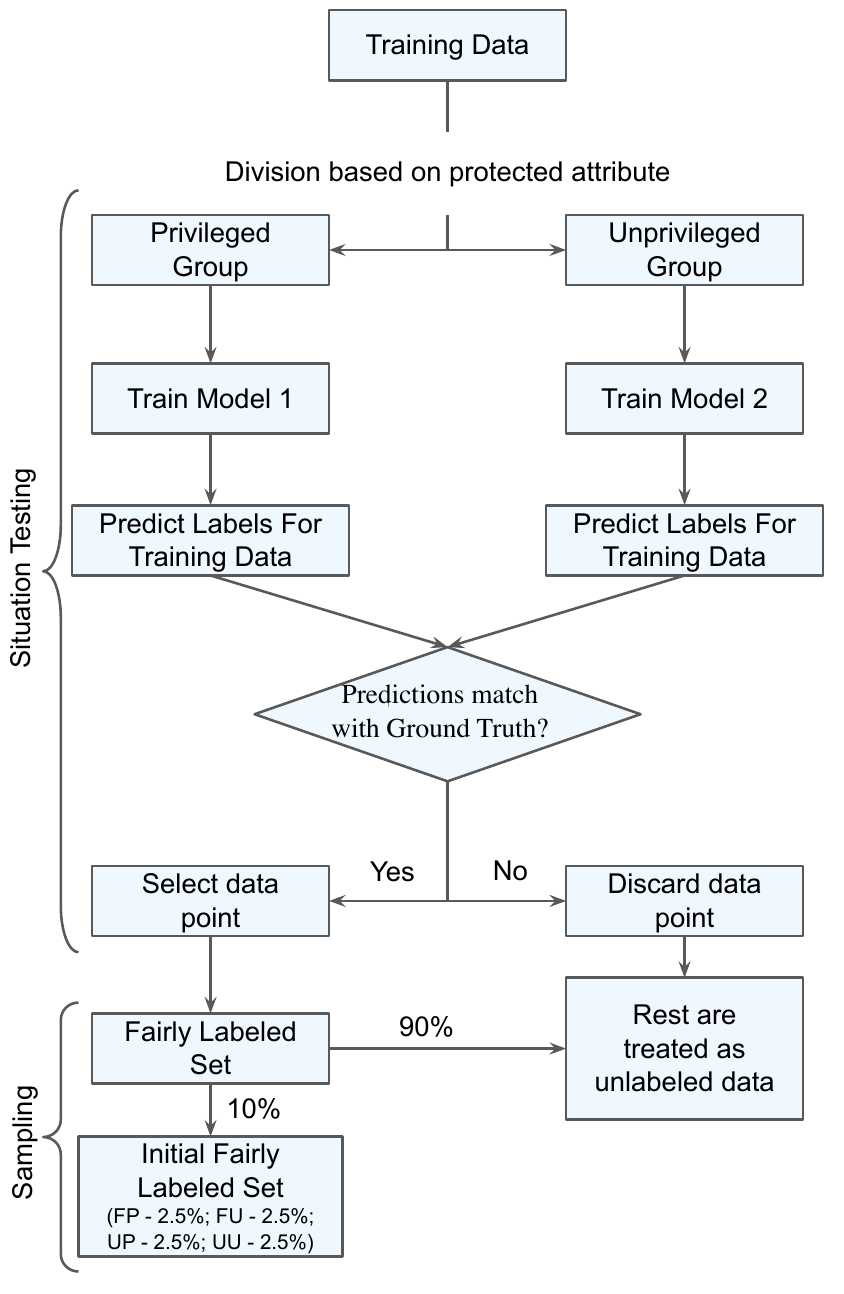}}
\caption{Algorithm for selecting/sampling fairly labeled data points from the available data. }
\label{Initial_selection}
\end{figure}

After this \textit{situation testing} phase, we get the \textit{fairly labeled set}. But we do not use this whole set for training. We take a small portion (10\%) of this set to start. This selection is not a random selection. The normal trend of semi-supervised learning is keeping the initial labeled data balanced based on class so that the semi-supervised model gets an opportunity to learn features from all the classes equally~\cite{hyun2020classimbalanced,10.5555/2283696.2283708}. In fairness, data balancing helps more if instead of just class, protected attributes are also balanced~\cite{chen2018classifier,10.1145/3340531.3411980,Chakraborty2021BiasIM}. Therefore, we select data points in a way that the \textit{initial fairly labeled set} has equal proportion of ``favorable-privileged (FP)'' (for ``Adult'' dataset - ``high income'' \& ``male''  ), ``favorable-unprivileged (FU)'' (``high income'' \& ``female''  ), ``unfavorable-privileged (UP)'' (``low income'' \& ``male''  ), \& ``unfavorable-unprivileged (UU)'' (``low income'' \& ``female''  ) samples. Figure \ref{Initial_selection} shows the block diagram of the combined process of situation testing and sampling that generates perfectly balanced \textit{initial fairly labeled set}. 

\vspace{-0.3cm}

\subsection{Pseudo Labeling the Unlabeled data}
\label{methodology_part_2}

We have the \textit{initial fairly labeled set} from the previous step. We remove the labels of the rest of the training data and treat it as \textit{unlabeled data}. In this step, we pseudo-label the unlabeled data using four different semi-supervised techniques~\cite{Zhu06semi-supervisedlearning}. We followed a single pattern while applying all four techniques. We start with 
labeled dataset $D_{l}$ (\textit{initial fairly labeled set}), and unlabeled dataset $D_{u}$. Our goal is to get a new training dataset (pseudo-labels added). To do that, we select a baseline model (based on the technique). The baseline model is trained on the \textit{initial fairly labeled set}. Then baseline model predicts labels of the unlabeled data. We consider prediction of only those 
data points as valid where the confidence of the predictor is very high. There are two kinds of selection criterion used - (a) select k\_best data points based on prediction probability, or (b) select data points where prediction probability is above a certain threshold. Based on scikit-learn semi-supervised article~\cite{sklearn_probability}, the ideal probability threshold is 0.7. We have used the same value. The data points being predicted with more than 70\% confidence along with their predicted labels are added to the training data. This process is repeated until max\_iteration is reached. Now we will describe four semi-supervised approaches in detail. \\
\textbf{Self Training}: Self-training~\cite{10.3115/981658.981684} requires a baseline model. We have used logistic regression because it returns well calibrated predictions as it directly optimizes \textit{log loss}~\cite{sklearn_calibration,log_linear_models}. At first, a supervised classifier (here logistic regression) is trained on the  \textit{initial fairly labeled set} and then incrementally unlabeled data points are predicted. At each iteration, the data points having prediction probability more than ``probability\_threshold'' (0.7) are selected and added to the training set with the predicted labels. This process continues until max\_iteration is reached. Finally, as a result, we get a new training dataset that contains \textit{initial fairly labeled set} and pseudo-labeled data points (by self-training). \\
\textbf{ Label Propagation}:
Label propagation is a semi-supervised graph inference algorithm~\cite{Zhu02learningfrom,LPADapeng}. The algorithm starts with building a graph from the available labeled and unlabeled data. Each data point is a node in the graph and edges are the similarity weights. The graph is represented in the form of a matrix. At first, a unique label is assigned to each node in the network. At time t = 0, for a node x, let its label is $C_{x}(0) = x$. The value of t is incremented. After that, for each node x in the network, the most frequently occurring label among all the nodes with which x is connected is found out. Finally, convergence condition is checked. If not met, we repeat, else stop.\\
\textbf{Label Spreading}:
Label spreading~\cite{NIPS2003_87682805} is also a graph inference algorithm but has some differences from label propagation. Label propagation uses the raw similarity matrix constructed from the data with no modifications. It believes that the original labels are perfect. On the contrary, label spreading does not blindly believe the original labels and makes modifications to the ground truth. The label spreading algorithm iterates on a modified version of the original graph and normalizes the edge weights by computing the normalized graph Laplacian matrix.\\
\textbf{Co-Training}:
Co-training is a very popular semi-supervised approach developed by Blum et al.~\cite{10.1145/279943.279962}. It is based on \textit{majority voting} technique. At first, separate models are built from each attribute of \textit{initial fairly labeled set}. Then each model predicts labels for the unlabeled data. For every data point, predictions for all the models are checked to get the majority voting. The data point is added to the train set with the majority label. Same procedure is incrementally repeated for all the unlabeled data points. Here also, we use logistic regression model (one model per feature) for good calibration. 

\begin{algorithm}[!t]
\footnotesize
\DontPrintSemicolon
    \KwInput{Dataset, Protected Attribute(p\_attrs), Class Label(cl)}
    \KwOutput{Balanced Dataset}
    \SetKwProg{Fn}{Def}{:}{}
      \Fn{\FSub{$Dataset$, $p\_attrs$, $cl$}}{
    count\_groups = get\_count(Dataset, p\_attrs, cl) \;
    % \tcc{get\_count method returns dict with size of sub groups in dataset divided by cl first and then p\_attrs}
    
    max\_size = max(count\_groups) \;
    % \tcp*{maximum size of sub groups}
    
    cr, f = 0.8, 0.8  (user can pick any value in [0,1]) \;
    % \tcp*{crossover frequency}

    % \tcp*{mutation amount}
    
    \For{c in cl}
    {
        \For{attr in p\_attrs}
        {
            sub\_data = Dataset(cl=c $\And$ p\_attrs=attr)\;
            sub\_group\_size = count\_groups[c][attr]\;
            to\_generate = max\_size - sub\_group\_size\;
            knn = NearestNeighbors(sub\_data)\;
            \For{i in range(to\_generate)}
            {
                parent = Dataset[rand\_sample\_id]\;
                ngbr = knn.kneighbors(parent, 2)\;
                c1, c2 = Dataset[ngbr[0]], Dataset[ngbr[1]]\;
                new\_candidate = []\;
                \For{col in parent.columns}
                {
                    \If{cr > random(0,1)}
                    {
                        new\_val = p[col] + f*(c1[col]-c2[col])\;
                    }
                    \Else
                    {
                        new\_val = p[col]\;
                    }
                    new\_candidate.add(new\_val)\;
                    
                }
            }
            Dataset.add(new\_candidate)
        }
    }
    return Dataset
}   
\caption{Oversampling pseudocode inspired from~\cite{Chakraborty2021BiasIM}}
\label{Algo_Fair_SMOTE}
\end{algorithm}

\begin{figure}[]
\centering
\adjustbox{max height=3in}{
\includegraphics[width=\linewidth]{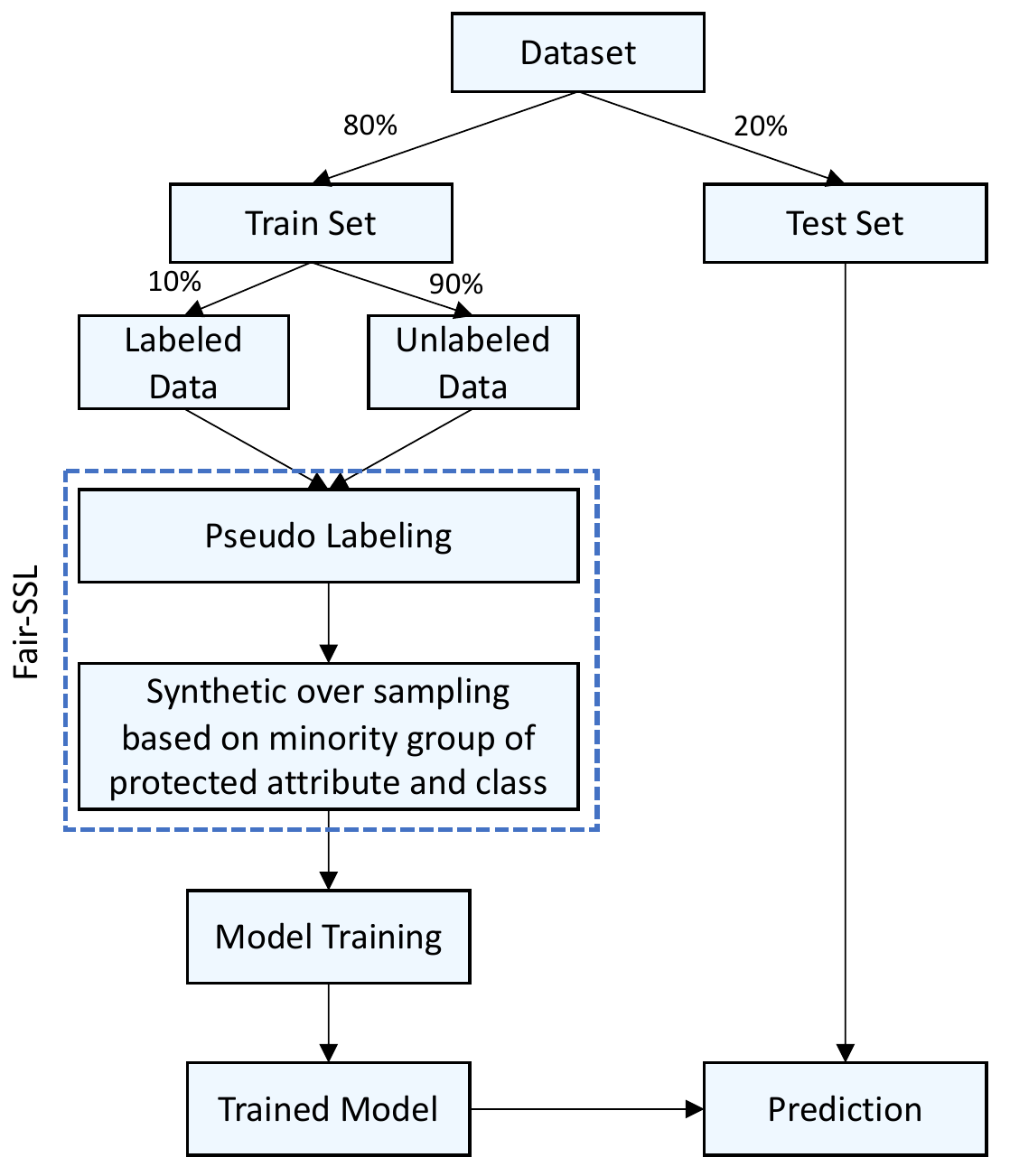}}
\caption{Framework of Fair-SSL}
\label{framework}
\end{figure}
\hspace{-2cm}

\subsection{Synthetic Oversampling \& Balancing}
\label{methodology_part_3}
Our training data is now labeled. We can go for model training. But some prior works~\cite{chen2018classifier,Chakraborty2021BiasIM} claimed that training data needs to be balanced in order to achieve fair prediction. Here data balancing means the number of examples based on protected attribute and class should be almost equal. We used the Fair-SMOTE technique (algorithm ~\ref{Algo_Fair_SMOTE}) by Chakraborty et al.~\cite{Chakraborty2021BiasIM} to achieve that. Initial training data are divided into four groups (Favorable \& Privileged, Favorable \& Unprivileged, Unfavorable \& Privileged, Unfavorable \& Unprivileged). Initially, these subgroups are of unequal sizes. After synthetic oversampling, the training data becomes balanced based on target class and protected attribute i.e. above mentioned four groups become of equal sizes. 
We have now described all the major components of Fair-SSL that generates pseudo-labeled, and balanced training data from a very small set of labeled examples. Figure \ref{framework} shows Fair-SSL framework. 

\section{Results}
\label{results}
We conducted our experiments on ten datasets (Table \ref{datasets}) and used three learners (logistic regression, random forest, support vector machine (svm)). Because of small size of datasets, prior fairness works~\cite{Chakraborty_2020,10.1007/978-3-642-33486-3_3,NIPS2017_6988,Biswas_2020,chakraborty2019software,Chakraborty2021BiasIM} also chose simple models like us instead of deep learning models. We split the datasets using 5 fold cross-validation (train - 80\%, test - 20\%) and repeat 10 times with random seeds and finally report the median for every experiment. We compared Fair-SSL with three state-of-the-art fairness algorithms - Optimized Pre-processing~\cite{NIPS2017_6988}, Fairway~\cite{Chakraborty_2020}, and  Fair-SMOTE~\cite{Chakraborty2021BiasIM} using Scott-Knott statistical test~\cite{mittas2013ranking, ghotra2015revisiting}.

\begin{table*}[]
\scriptsize
\caption{RQ1, RQ2, and RQ4 results: ``Default'' means off-the-shelf logistic regression; Optimized Pre-processing(OP)~\cite{NIPS2017_6988}, Fairway~\cite{Chakraborty_2020}, \& Fair-SMOTE~\cite{Chakraborty2021BiasIM} are state-of-the-art supervised methods. ST is self-training; LP is label-propagation; LS is label-spreading; CT is co-training. \colorbox{deeppink}{Darkest} = 1st Rank;\colorbox{pink}{Lighter} = 2nd Rank; \colorbox{lightpink}{Lightest} = 3rd Rank; White = Last Rank (note: for the metrics with `+' more is better and for the metrics with `-' less is better). Rankings were calculated via Scott-Knott test.}
\label{RQ2_3}
\begin{tabular}{cccccccc
>{\columncolor[HTML]{D28986}}c 
>{\columncolor[HTML]{D28986}}c 
>{\columncolor[HTML]{D28986}}c 
>{\columncolor[HTML]{D28986}}c }
\cellcolor[HTML]{C0C0C0}Dataset                                                                        & \cellcolor[HTML]{C0C0C0}\begin{tabular}[c]{@{}c@{}}Protected\\ Attribute\end{tabular} & \cellcolor[HTML]{C0C0C0}Algorithms & \cellcolor[HTML]{C0C0C0}\begin{tabular}[c]{@{}c@{}}Recall\\ (+)\end{tabular}      & \cellcolor[HTML]{C0C0C0}\begin{tabular}[c]{@{}c@{}}False alarm\\ (-)\end{tabular} & \cellcolor[HTML]{C0C0C0}\begin{tabular}[c]{@{}c@{}}Precision\\ (+)\end{tabular}   & \cellcolor[HTML]{C0C0C0}\begin{tabular}[c]{@{}c@{}}Accuracy\\ (+)\end{tabular}    & \cellcolor[HTML]{C0C0C0}\begin{tabular}[c]{@{}c@{}}F1 Score\\ (+)\end{tabular}    & \cellcolor[HTML]{C0C0C0}\begin{tabular}[c]{@{}c@{}}AOD\\ (-)\end{tabular}         & \cellcolor[HTML]{C0C0C0}\begin{tabular}[c]{@{}c@{}}EOD\\ (-)\end{tabular}         & \cellcolor[HTML]{C0C0C0}\begin{tabular}[c]{@{}c@{}}SPD\\ (-)\end{tabular}         & \cellcolor[HTML]{C0C0C0}\begin{tabular}[c]{@{}c@{}}DI\\ (-)\end{tabular}          \\ \hline
\multicolumn{1}{|c|}{}                                                                                 & \multicolumn{1}{c|}{}                                                                 & \multicolumn{1}{c|}{Default}       & \multicolumn{1}{c|}{\cellcolor[HTML]{FFABA7}{\color[HTML]{333333} \textbf{0.46}}} & \multicolumn{1}{c|}{\cellcolor[HTML]{FFABA7}{\color[HTML]{333333} \textbf{0.07}}} & \multicolumn{1}{c|}{\cellcolor[HTML]{D28986}{\color[HTML]{333333} \textbf{0.69}}} & \multicolumn{1}{c|}{\cellcolor[HTML]{D28986}{\color[HTML]{333333} \textbf{0.83}}} & \multicolumn{1}{c|}{\cellcolor[HTML]{FFABA7}{\color[HTML]{333333} \textbf{0.54}}} & \multicolumn{1}{c|}{\cellcolor[HTML]{FFFFFF}{\color[HTML]{333333} \textbf{0.12}}} & \multicolumn{1}{c|}{\cellcolor[HTML]{FFFFFF}{\color[HTML]{333333} \textbf{0.24}}} & \multicolumn{1}{c|}{\cellcolor[HTML]{FFFFFF}{\color[HTML]{333333} \textbf{0.21}}} & \multicolumn{1}{c|}{\cellcolor[HTML]{FFFFFF}{\color[HTML]{333333} \textbf{0.56}}} \\ \cline{3-12} 
\multicolumn{1}{|c|}{}                                                                                 & \multicolumn{1}{c|}{}                                                                 & \multicolumn{1}{c|}{OP}            & \multicolumn{1}{c|}{\cellcolor[HTML]{FFABA7}\textbf{0.42}}                        & \multicolumn{1}{c|}{\cellcolor[HTML]{FFABA7}\textbf{0.09}}                        & \multicolumn{1}{c|}{\cellcolor[HTML]{FFABA7}\textbf{0.61}}                        & \multicolumn{1}{c|}{\cellcolor[HTML]{FFABA7}\textbf{0.76}}                        & \multicolumn{1}{c|}{\cellcolor[HTML]{FFCCC9}\textbf{0.51}}                        & \multicolumn{1}{c|}{\cellcolor[HTML]{D28986}\textbf{0.04}}                        & \multicolumn{1}{c|}{\cellcolor[HTML]{D28986}\textbf{0.03}}                        & \multicolumn{1}{c|}{\cellcolor[HTML]{D28986}\textbf{0.04}}                        & \multicolumn{1}{c|}{\cellcolor[HTML]{FFABA7}\textbf{0.14}}                        \\ \cline{3-12} 
\multicolumn{1}{|c|}{}                                                                                 & \multicolumn{1}{c|}{}                                                                 & \multicolumn{1}{c|}{Fairway}       & \multicolumn{1}{c|}{\textbf{0.25}}                                                & \multicolumn{1}{c|}{\cellcolor[HTML]{D28986}\textbf{0.04}}                        & \multicolumn{1}{c|}{\cellcolor[HTML]{D28986}\textbf{0.71}}                        & \multicolumn{1}{c|}{\cellcolor[HTML]{FFCCC9}\textbf{0.72}}                        & \multicolumn{1}{c|}{\textbf{0.42}}                                                & \multicolumn{1}{c|}{\cellcolor[HTML]{D28986}\textbf{0.02}}                        & \multicolumn{1}{c|}{\cellcolor[HTML]{D28986}\textbf{0.03}}                        & \multicolumn{1}{c|}{\cellcolor[HTML]{D28986}\textbf{0.04}}                        & \multicolumn{1}{c|}{\cellcolor[HTML]{FFABA7}\textbf{0.11}}                        \\ \cline{3-12} 
\multicolumn{1}{|c|}{}                                                                                 & \multicolumn{1}{c|}{}                                                                 & \multicolumn{1}{c|}{Fair-SMOTE}    & \multicolumn{1}{c|}{\cellcolor[HTML]{D28986}{\color[HTML]{333333} \textbf{0.71}}} & \multicolumn{1}{c|}{\cellcolor[HTML]{FFCCC9}{\color[HTML]{333333} \textbf{0.25}}} & \multicolumn{1}{c|}{\cellcolor[HTML]{FFCCC9}{\color[HTML]{333333} \textbf{0.51}}} & \multicolumn{1}{c|}{\cellcolor[HTML]{FFCCC9}{\color[HTML]{333333} \textbf{0.73}}} & \multicolumn{1}{c|}{\cellcolor[HTML]{D28986}{\color[HTML]{333333} \textbf{0.62}}} & \multicolumn{1}{c|}{\cellcolor[HTML]{D28986}{\color[HTML]{333333} \textbf{0.01}}} & \multicolumn{1}{c|}{\cellcolor[HTML]{D28986}{\color[HTML]{333333} \textbf{0.02}}} & \multicolumn{1}{c|}{\cellcolor[HTML]{D28986}{\color[HTML]{333333} \textbf{0.03}}} & \multicolumn{1}{c|}{\cellcolor[HTML]{D28986}{\color[HTML]{333333} \textbf{0.08}}} \\ \cline{3-12} 
\multicolumn{1}{|c|}{}                                                                                 & \multicolumn{1}{c|}{}                                                                 & \multicolumn{1}{c|}{Fair-SSL-ST}   & \multicolumn{1}{c|}{\cellcolor[HTML]{D28986}{\color[HTML]{333333} \textbf{0.71}}} & \multicolumn{1}{c|}{\cellcolor[HTML]{FFFFFF}{\color[HTML]{333333} \textbf{0.39}}} & \multicolumn{1}{c|}{\cellcolor[HTML]{FFFFFF}{\color[HTML]{333333} \textbf{0.42}}} & \multicolumn{1}{c|}{\cellcolor[HTML]{FFFFFF}{\color[HTML]{333333} \textbf{0.62}}} & \multicolumn{1}{c|}{\cellcolor[HTML]{FFCCC9}{\color[HTML]{333333} \textbf{0.51}}} & \multicolumn{1}{c|}{\cellcolor[HTML]{D28986}{\color[HTML]{333333} \textbf{0.04}}} & \multicolumn{1}{c|}{\cellcolor[HTML]{D28986}{\color[HTML]{333333} \textbf{0.03}}} & \multicolumn{1}{c|}{\cellcolor[HTML]{D28986}{\color[HTML]{333333} \textbf{0.07}}} & \multicolumn{1}{c|}{\cellcolor[HTML]{FFABA7}{\color[HTML]{333333} \textbf{0.12}}} \\ \cline{3-12} 
\multicolumn{1}{|c|}{}                                                                                 & \multicolumn{1}{c|}{}                                                                 & \multicolumn{1}{c|}{Fair-SSL-LP}   & \multicolumn{1}{c|}{\cellcolor[HTML]{D28986}{\color[HTML]{333333} \textbf{0.72}}} & \multicolumn{1}{c|}{\cellcolor[HTML]{FFFFFF}{\color[HTML]{333333} \textbf{0.38}}} & \multicolumn{1}{c|}{\cellcolor[HTML]{FFFFFF}{\color[HTML]{333333} \textbf{0.45}}} & \multicolumn{1}{c|}{\cellcolor[HTML]{FFFFFF}{\color[HTML]{333333} \textbf{0.66}}} & \multicolumn{1}{c|}{\cellcolor[HTML]{FFCCC9}{\color[HTML]{333333} \textbf{0.53}}} & \multicolumn{1}{c|}{\cellcolor[HTML]{D28986}{\color[HTML]{333333} \textbf{0.03}}} & \multicolumn{1}{c|}{\cellcolor[HTML]{D28986}{\color[HTML]{333333} \textbf{0.03}}} & \multicolumn{1}{c|}{\cellcolor[HTML]{D28986}{\color[HTML]{333333} \textbf{0.04}}} & \multicolumn{1}{c|}{\cellcolor[HTML]{D28986}{\color[HTML]{333333} \textbf{0.05}}} \\ \cline{3-12} 
\multicolumn{1}{|c|}{}                                                                                 & \multicolumn{1}{c|}{}                                                                 & \multicolumn{1}{c|}{Fair-SSL-LS}   & \multicolumn{1}{c|}{\cellcolor[HTML]{D28986}{\color[HTML]{333333} \textbf{0.72}}} & \multicolumn{1}{c|}{\cellcolor[HTML]{FFFFFF}{\color[HTML]{333333} \textbf{0.31}}} & \multicolumn{1}{c|}{\cellcolor[HTML]{FFFFFF}{\color[HTML]{333333} \textbf{0.42}}} & \multicolumn{1}{c|}{\cellcolor[HTML]{FFCCC9}{\color[HTML]{333333} \textbf{0.71}}} & \multicolumn{1}{c|}{\cellcolor[HTML]{FFABA7}{\color[HTML]{333333} \textbf{0.55}}} & \multicolumn{1}{c|}{\cellcolor[HTML]{D28986}{\color[HTML]{333333} \textbf{0.03}}} & \multicolumn{1}{c|}{\cellcolor[HTML]{D28986}{\color[HTML]{333333} \textbf{0.04}}} & \multicolumn{1}{c|}{\cellcolor[HTML]{D28986}{\color[HTML]{333333} \textbf{0.06}}} & \multicolumn{1}{c|}{\cellcolor[HTML]{D28986}{\color[HTML]{333333} \textbf{0.08}}} \\ \cline{3-12} 
\multicolumn{1}{|c|}{\multirow{-8}{*}{\begin{tabular}[c]{@{}c@{}}Adult Census \\ Income\end{tabular}}} & \multicolumn{1}{c|}{\multirow{-8}{*}{Sex}}                                            & \multicolumn{1}{c|}{Fair-SSL-CT}   & \multicolumn{1}{c|}{\cellcolor[HTML]{D28986}{\color[HTML]{333333} \textbf{0.76}}} & \multicolumn{1}{c|}{\cellcolor[HTML]{FFFFFF}{\color[HTML]{333333} \textbf{0.35}}} & \multicolumn{1}{c|}{\cellcolor[HTML]{FFFFFF}{\color[HTML]{333333} \textbf{0.44}}} & \multicolumn{1}{c|}{\cellcolor[HTML]{FFFFFF}{\color[HTML]{333333} \textbf{0.68}}} & \multicolumn{1}{c|}{\cellcolor[HTML]{FFABA7}{\color[HTML]{333333} \textbf{0.57}}} & \multicolumn{1}{c|}{\cellcolor[HTML]{D28986}{\color[HTML]{333333} \textbf{0.06}}} & \multicolumn{1}{c|}{\cellcolor[HTML]{D28986}{\color[HTML]{333333} \textbf{0.04}}} & \multicolumn{1}{c|}{\cellcolor[HTML]{D28986}{\color[HTML]{333333} \textbf{0.03}}} & \multicolumn{1}{c|}{\cellcolor[HTML]{D28986}{\color[HTML]{333333} \textbf{0.08}}} \\ \hline
\multicolumn{1}{|c|}{}                                                                                 & \multicolumn{1}{c|}{}                                                                 & \multicolumn{1}{c|}{Default}       & \multicolumn{1}{c|}{\cellcolor[HTML]{D28986}{\color[HTML]{333333} \textbf{0.67}}} & \multicolumn{1}{c|}{\cellcolor[HTML]{FFFFFF}{\color[HTML]{333333} \textbf{0.38}}} & \multicolumn{1}{c|}{\cellcolor[HTML]{D28986}{\color[HTML]{333333} \textbf{0.66}}} & \multicolumn{1}{c|}{\cellcolor[HTML]{D28986}{\color[HTML]{333333} \textbf{0.64}}} & \multicolumn{1}{c|}{\cellcolor[HTML]{FFABA7}{\color[HTML]{333333} \textbf{0.61}}} & \multicolumn{1}{c|}{\cellcolor[HTML]{FFFFFF}{\color[HTML]{333333} \textbf{0.09}}} & \multicolumn{1}{c|}{\cellcolor[HTML]{FFFFFF}{\color[HTML]{333333} \textbf{0.19}}} & \multicolumn{1}{c|}{\cellcolor[HTML]{FFFFFF}{\color[HTML]{333333} \textbf{0.18}}} & \multicolumn{1}{c|}{\cellcolor[HTML]{FFFFFF}{\color[HTML]{333333} \textbf{0.28}}} \\ \cline{3-12} 
\multicolumn{1}{|c|}{}                                                                                 & \multicolumn{1}{c|}{}                                                                 & \multicolumn{1}{c|}{OP}            & \multicolumn{1}{c|}{\cellcolor[HTML]{D28986}\textbf{0.71}}                        & \multicolumn{1}{c|}{\textbf{0.36}}                                                & \multicolumn{1}{c|}{\cellcolor[HTML]{D28986}\textbf{0.64}}                        & \multicolumn{1}{c|}{\cellcolor[HTML]{FFABA7}\textbf{0.62}}                        & \multicolumn{1}{c|}{\cellcolor[HTML]{FFABA7}\textbf{0.60}}                        & \multicolumn{1}{c|}{\cellcolor[HTML]{D28986}\textbf{0.04}}                        & \multicolumn{1}{c|}{\cellcolor[HTML]{D28986}\textbf{0.03}}                        & \multicolumn{1}{c|}{\cellcolor[HTML]{D28986}\textbf{0.05}}                        & \multicolumn{1}{c|}{\cellcolor[HTML]{D28986}\textbf{0.08}}                        \\ \cline{3-12} 
\multicolumn{1}{|c|}{}                                                                                 & \multicolumn{1}{c|}{}                                                                 & \multicolumn{1}{c|}{Fairway}       & \multicolumn{1}{c|}{\cellcolor[HTML]{FFCCC9}\textbf{0.56}}                        & \multicolumn{1}{c|}{\cellcolor[HTML]{D28986}\textbf{0.22}}                        & \multicolumn{1}{c|}{\cellcolor[HTML]{FFFFFF}\textbf{0.57}}                        & \multicolumn{1}{c|}{\cellcolor[HTML]{FFCCC9}\textbf{0.58}}                        & \multicolumn{1}{c|}{\cellcolor[HTML]{FFCCC9}\textbf{0.58}}                        & \multicolumn{1}{c|}{\cellcolor[HTML]{D28986}\textbf{0.03}}                        & \multicolumn{1}{c|}{\cellcolor[HTML]{D28986}\textbf{0.03}}                        & \multicolumn{1}{c|}{\cellcolor[HTML]{D28986}\textbf{0.06}}                        & \multicolumn{1}{c|}{\cellcolor[HTML]{D28986}\textbf{0.08}}                        \\ \cline{3-12} 
\multicolumn{1}{|c|}{}                                                                                 & \multicolumn{1}{c|}{}                                                                 & \multicolumn{1}{c|}{Fair-SMOTE}    & \multicolumn{1}{c|}{\cellcolor[HTML]{FFABA7}{\color[HTML]{333333} \textbf{0.62}}} & \multicolumn{1}{c|}{\cellcolor[HTML]{FFABA7}{\color[HTML]{333333} \textbf{0.32}}} & \multicolumn{1}{c|}{\cellcolor[HTML]{FFFFFF}{\color[HTML]{333333} \textbf{0.56}}} & \multicolumn{1}{c|}{\cellcolor[HTML]{FFFFFF}{\color[HTML]{333333} \textbf{0.55}}} & \multicolumn{1}{c|}{\cellcolor[HTML]{D28986}{\color[HTML]{333333} \textbf{0.65}}} & \multicolumn{1}{c|}{\cellcolor[HTML]{D28986}{\color[HTML]{333333} \textbf{0.02}}} & \multicolumn{1}{c|}{\cellcolor[HTML]{D28986}{\color[HTML]{333333} \textbf{0.05}}} & \multicolumn{1}{c|}{\cellcolor[HTML]{D28986}{\color[HTML]{333333} \textbf{0.08}}} & \multicolumn{1}{c|}{\cellcolor[HTML]{D28986}{\color[HTML]{333333} \textbf{0.09}}} \\ \cline{3-12} 
\multicolumn{1}{|c|}{}                                                                                 & \multicolumn{1}{c|}{}                                                                 & \multicolumn{1}{c|}{Fair-SSL-ST}   & \multicolumn{1}{c|}{\cellcolor[HTML]{FFFFFF}{\color[HTML]{333333} \textbf{0.42}}} & \multicolumn{1}{c|}{\cellcolor[HTML]{D28986}{\color[HTML]{333333} \textbf{0.21}}} & \multicolumn{1}{c|}{\cellcolor[HTML]{D28986}{\color[HTML]{333333} \textbf{0.65}}} & \multicolumn{1}{c|}{\cellcolor[HTML]{FFCCC9}{\color[HTML]{333333} \textbf{0.58}}} & \multicolumn{1}{c|}{\cellcolor[HTML]{FFFFFF}{\color[HTML]{333333} \textbf{0.54}}} & \multicolumn{1}{c|}{\cellcolor[HTML]{D28986}{\color[HTML]{333333} \textbf{0.02}}} & \multicolumn{1}{c|}{\cellcolor[HTML]{D28986}{\color[HTML]{333333} \textbf{0.09}}} & \multicolumn{1}{c|}{\cellcolor[HTML]{D28986}{\color[HTML]{333333} \textbf{0.12}}} & \multicolumn{1}{c|}{\cellcolor[HTML]{FFABA7}{\color[HTML]{333333} \textbf{0.21}}} \\ \cline{3-12} 
\multicolumn{1}{|c|}{}                                                                                 & \multicolumn{1}{c|}{}                                                                 & \multicolumn{1}{c|}{Fair-SSL-LP}   & \multicolumn{1}{c|}{\cellcolor[HTML]{FFFFFF}{\color[HTML]{333333} \textbf{0.52}}} & \multicolumn{1}{c|}{\cellcolor[HTML]{FFCCC9}{\color[HTML]{333333} \textbf{0.34}}} & \multicolumn{1}{c|}{\cellcolor[HTML]{FFABA7}{\color[HTML]{333333} \textbf{0.62}}} & \multicolumn{1}{c|}{\cellcolor[HTML]{FFFFFF}{\color[HTML]{333333} \textbf{0.54}}} & \multicolumn{1}{c|}{\cellcolor[HTML]{FFCCC9}{\color[HTML]{333333} \textbf{0.58}}} & \multicolumn{1}{c|}{\cellcolor[HTML]{D28986}{\color[HTML]{333333} \textbf{0.02}}} & \multicolumn{1}{c|}{\cellcolor[HTML]{D28986}{\color[HTML]{333333} \textbf{0.06}}} & \multicolumn{1}{c|}{\cellcolor[HTML]{D28986}{\color[HTML]{333333} \textbf{0.09}}} & \multicolumn{1}{c|}{\cellcolor[HTML]{FFABA7}{\color[HTML]{333333} \textbf{0.19}}} \\ \cline{3-12} 
\multicolumn{1}{|c|}{}                                                                                 & \multicolumn{1}{c|}{}                                                                 & \multicolumn{1}{c|}{Fair-SSL-LS}   & \multicolumn{1}{c|}{\cellcolor[HTML]{FFFFFF}{\color[HTML]{333333} \textbf{0.52}}} & \multicolumn{1}{c|}{\cellcolor[HTML]{FFCCC9}{\color[HTML]{333333} \textbf{0.33}}} & \multicolumn{1}{c|}{\cellcolor[HTML]{FFABA7}{\color[HTML]{333333} \textbf{0.61}}} & \multicolumn{1}{c|}{\cellcolor[HTML]{FFABA7}{\color[HTML]{333333} \textbf{0.62}}} & \multicolumn{1}{c|}{\cellcolor[HTML]{FFABA7}{\color[HTML]{333333} \textbf{0.62}}} & \multicolumn{1}{c|}{\cellcolor[HTML]{D28986}{\color[HTML]{333333} \textbf{0.03}}} & \multicolumn{1}{c|}{\cellcolor[HTML]{D28986}{\color[HTML]{333333} \textbf{0.03}}} & \multicolumn{1}{c|}{\cellcolor[HTML]{D28986}{\color[HTML]{333333} \textbf{0.05}}} & \multicolumn{1}{c|}{\cellcolor[HTML]{D28986}{\color[HTML]{333333} \textbf{0.09}}} \\ \cline{3-12} 
\multicolumn{1}{|c|}{\multirow{-8}{*}{Compas}}                                                         & \multicolumn{1}{c|}{\multirow{-8}{*}{Sex}}                                            & \multicolumn{1}{c|}{Fair-SSL-CT}   & \multicolumn{1}{c|}{\cellcolor[HTML]{FFFFFF}{\color[HTML]{333333} \textbf{0.49}}} & \multicolumn{1}{c|}{\cellcolor[HTML]{FFABA7}{\color[HTML]{333333} \textbf{0.28}}} & \multicolumn{1}{c|}{\cellcolor[HTML]{FFABA7}{\color[HTML]{333333} \textbf{0.62}}} & \multicolumn{1}{c|}{\cellcolor[HTML]{FFABA7}{\color[HTML]{333333} \textbf{0.61}}} & \multicolumn{1}{c|}{\cellcolor[HTML]{FFCCC9}{\color[HTML]{333333} \textbf{0.57}}} & \multicolumn{1}{c|}{\cellcolor[HTML]{D28986}{\color[HTML]{333333} \textbf{0.03}}} & \multicolumn{1}{c|}{\cellcolor[HTML]{D28986}{\color[HTML]{333333} \textbf{0.02}}} & \multicolumn{1}{c|}{\cellcolor[HTML]{D28986}{\color[HTML]{333333} \textbf{0.05}}} & \multicolumn{1}{c|}{\cellcolor[HTML]{D28986}{\color[HTML]{333333} \textbf{0.07}}} \\ \hline
\multicolumn{1}{|c|}{}                                                                                 & \multicolumn{1}{c|}{}                                                                 & \multicolumn{1}{c|}{Default}       & \multicolumn{1}{c|}{\cellcolor[HTML]{FFFFFF}{\color[HTML]{333333} \textbf{0.25}}} & \multicolumn{1}{c|}{\cellcolor[HTML]{D28986}{\color[HTML]{333333} \textbf{0.07}}} & \multicolumn{1}{c|}{\cellcolor[HTML]{D28986}{\color[HTML]{333333} \textbf{0.70}}} & \multicolumn{1}{c|}{\cellcolor[HTML]{D28986}{\color[HTML]{333333} \textbf{0.78}}} & \multicolumn{1}{c|}{\cellcolor[HTML]{FFFFFF}{\color[HTML]{333333} \textbf{0.34}}} & \multicolumn{1}{c|}{\cellcolor[HTML]{FFFFFF}{\color[HTML]{333333} \textbf{0.05}}} & \multicolumn{1}{c|}{\cellcolor[HTML]{FFFFFF}{\color[HTML]{333333} \textbf{0.08}}} & \multicolumn{1}{c|}{\cellcolor[HTML]{FFFFFF}{\color[HTML]{333333} \textbf{0.06}}} & \multicolumn{1}{c|}{\cellcolor[HTML]{FFFFFF}{\color[HTML]{333333} \textbf{0.35}}} \\ \cline{3-12} 
\multicolumn{1}{|c|}{}                                                                                 & \multicolumn{1}{c|}{}                                                                 & \multicolumn{1}{c|}{OP}            & \multicolumn{1}{c|}{\cellcolor[HTML]{FFCCC9}\textbf{0.28}}                        & \multicolumn{1}{c|}{\cellcolor[HTML]{D28986}\textbf{0.06}}                        & \multicolumn{1}{c|}{\cellcolor[HTML]{FFABA7}\textbf{0.65}}                        & \multicolumn{1}{c|}{\cellcolor[HTML]{FFABA7}\textbf{0.70}}                        & \multicolumn{1}{c|}{\textbf{0.32}}                                                & \multicolumn{1}{c|}{\cellcolor[HTML]{D28986}\textbf{0.01}}                        & \multicolumn{1}{c|}{\cellcolor[HTML]{D28986}\textbf{0.02}}                        & \multicolumn{1}{c|}{\cellcolor[HTML]{D28986}\textbf{0.03}}                        & \multicolumn{1}{c|}{\cellcolor[HTML]{D28986}\textbf{0.09}}                        \\ \cline{3-12} 
\multicolumn{1}{|c|}{}                                                                                 & \multicolumn{1}{c|}{}                                                                 & \multicolumn{1}{c|}{Fairway}       & \multicolumn{1}{c|}{\textbf{0.21}}                                                & \multicolumn{1}{c|}{\cellcolor[HTML]{D28986}\textbf{0.04}}                        & \multicolumn{1}{c|}{\cellcolor[HTML]{FFABA7}\textbf{0.67}}                        & \multicolumn{1}{c|}{\cellcolor[HTML]{FFABA7}\textbf{0.67}}                        & \multicolumn{1}{c|}{\textbf{0.33}}                                                & \multicolumn{1}{c|}{\cellcolor[HTML]{D28986}\textbf{0.01}}                        & \multicolumn{1}{c|}{\cellcolor[HTML]{D28986}\textbf{0.04}}                        & \multicolumn{1}{c|}{\cellcolor[HTML]{D28986}\textbf{0.03}}                        & \multicolumn{1}{c|}{\cellcolor[HTML]{FFABA7}\textbf{0.12}}                        \\ \cline{3-12} 
\multicolumn{1}{|c|}{}                                                                                 & \multicolumn{1}{c|}{}                                                                 & \multicolumn{1}{c|}{Fair-SMOTE}    & \multicolumn{1}{c|}{\cellcolor[HTML]{D28986}{\color[HTML]{333333} \textbf{0.58}}} & \multicolumn{1}{c|}{\cellcolor[HTML]{FFCCC9}{\color[HTML]{333333} \textbf{0.26}}} & \multicolumn{1}{c|}{\cellcolor[HTML]{FFFFFF}{\color[HTML]{333333} \textbf{0.39}}} & \multicolumn{1}{c|}{\cellcolor[HTML]{FFABA7}{\color[HTML]{333333} \textbf{0.68}}} & \multicolumn{1}{c|}{\cellcolor[HTML]{FFABA7}{\color[HTML]{333333} \textbf{0.44}}} & \multicolumn{1}{c|}{\cellcolor[HTML]{D28986}{\color[HTML]{333333} \textbf{0.02}}} & \multicolumn{1}{c|}{\cellcolor[HTML]{D28986}{\color[HTML]{333333} \textbf{0.03}}} & \multicolumn{1}{c|}{\cellcolor[HTML]{D28986}{\color[HTML]{333333} \textbf{0.05}}} & \multicolumn{1}{c|}{\cellcolor[HTML]{D28986}{\color[HTML]{333333} \textbf{0.03}}} \\ \cline{3-12} 
\multicolumn{1}{|c|}{}                                                                                 & \multicolumn{1}{c|}{}                                                                 & \multicolumn{1}{c|}{Fair-SSL-ST}   & \multicolumn{1}{c|}{\cellcolor[HTML]{FFABA7}{\color[HTML]{333333} \textbf{0.48}}} & \multicolumn{1}{c|}{\cellcolor[HTML]{FFABA7}{\color[HTML]{333333} \textbf{0.12}}} & \multicolumn{1}{c|}{\cellcolor[HTML]{FFCCC9}{\color[HTML]{333333} \textbf{0.53}}} & \multicolumn{1}{c|}{\cellcolor[HTML]{D28986}{\color[HTML]{333333} \textbf{0.79}}} & \multicolumn{1}{c|}{\cellcolor[HTML]{D28986}{\color[HTML]{333333} \textbf{0.51}}} & \multicolumn{1}{c|}{\cellcolor[HTML]{D28986}{\color[HTML]{333333} \textbf{0.03}}} & \multicolumn{1}{c|}{\cellcolor[HTML]{D28986}{\color[HTML]{333333} \textbf{0.04}}} & \multicolumn{1}{c|}{\cellcolor[HTML]{D28986}{\color[HTML]{333333} \textbf{0.05}}} & \multicolumn{1}{c|}{\cellcolor[HTML]{FFABA7}{\color[HTML]{333333} \textbf{0.12}}} \\ \cline{3-12} 
\multicolumn{1}{|c|}{}                                                                                 & \multicolumn{1}{c|}{}                                                                 & \multicolumn{1}{c|}{Fair-SSL-LP}   & \multicolumn{1}{c|}{\cellcolor[HTML]{FFABA7}{\color[HTML]{333333} \textbf{0.51}}} & \multicolumn{1}{c|}{\cellcolor[HTML]{FFFFFF}{\color[HTML]{333333} \textbf{0.31}}} & \multicolumn{1}{c|}{\cellcolor[HTML]{FFFFFF}{\color[HTML]{333333} \textbf{0.37}}} & \multicolumn{1}{c|}{\cellcolor[HTML]{FFABA7}{\color[HTML]{333333} \textbf{0.67}}} & \multicolumn{1}{c|}{\cellcolor[HTML]{FFABA7}{\color[HTML]{333333} \textbf{0.44}}} & \multicolumn{1}{c|}{\cellcolor[HTML]{D28986}{\color[HTML]{333333} \textbf{0.05}}} & \multicolumn{1}{c|}{\cellcolor[HTML]{D28986}{\color[HTML]{333333} \textbf{0.03}}} & \multicolumn{1}{c|}{\cellcolor[HTML]{D28986}{\color[HTML]{333333} \textbf{0.03}}} & \multicolumn{1}{c|}{\cellcolor[HTML]{D28986}{\color[HTML]{333333} \textbf{0.08}}} \\ \cline{3-12} 
\multicolumn{1}{|c|}{}                                                                                 & \multicolumn{1}{c|}{}                                                                 & \multicolumn{1}{c|}{Fair-SSL-LS}   & \multicolumn{1}{c|}{\cellcolor[HTML]{D28986}{\color[HTML]{333333} \textbf{0.59}}} & \multicolumn{1}{c|}{\cellcolor[HTML]{FFFFFF}{\color[HTML]{333333} \textbf{0.36}}} & \multicolumn{1}{c|}{\cellcolor[HTML]{FFFFFF}{\color[HTML]{333333} \textbf{0.34}}} & \multicolumn{1}{c|}{\cellcolor[HTML]{FFFFFF}{\color[HTML]{333333} \textbf{0.64}}} & \multicolumn{1}{c|}{\cellcolor[HTML]{FFABA7}{\color[HTML]{333333} \textbf{0.42}}} & \multicolumn{1}{c|}{\cellcolor[HTML]{D28986}{\color[HTML]{333333} \textbf{0.05}}} & \multicolumn{1}{c|}{\cellcolor[HTML]{D28986}{\color[HTML]{333333} \textbf{0.04}}} & \multicolumn{1}{c|}{\cellcolor[HTML]{D28986}{\color[HTML]{333333} \textbf{0.03}}} & \multicolumn{1}{c|}{\cellcolor[HTML]{D28986}{\color[HTML]{333333} \textbf{0.09}}} \\ \cline{3-12} 
\multicolumn{1}{|c|}{\multirow{-8}{*}{\begin{tabular}[c]{@{}c@{}}Default \\ Credit\end{tabular}}}      & \multicolumn{1}{c|}{\multirow{-8}{*}{Sex}}                                            & \multicolumn{1}{c|}{Fair-SSL-CT}   & \multicolumn{1}{c|}{\cellcolor[HTML]{D28986}\textbf{0.59}}                        & \multicolumn{1}{c|}{\cellcolor[HTML]{FFFFFF}\textbf{0.35}}                        & \multicolumn{1}{c|}{\cellcolor[HTML]{FFFFFF}\textbf{0.36}}                        & \multicolumn{1}{c|}{\cellcolor[HTML]{FFFFFF}\textbf{0.65}}                        & \multicolumn{1}{c|}{\cellcolor[HTML]{FFABA7}\textbf{0.44}}                        & \multicolumn{1}{c|}{\cellcolor[HTML]{D28986}\textbf{0.03}}                        & \multicolumn{1}{c|}{\cellcolor[HTML]{D28986}\textbf{0.03}}                        & \multicolumn{1}{c|}{\cellcolor[HTML]{D28986}\textbf{0.05}}                        & \multicolumn{1}{c|}{\cellcolor[HTML]{D28986}\textbf{0.08}}                        \\ \hline
\multicolumn{1}{|c|}{}                                                                                 & \multicolumn{1}{c|}{}                                                                 & \multicolumn{1}{c|}{Default}       & \multicolumn{1}{c|}{\cellcolor[HTML]{FFABA7}{\color[HTML]{333333} \textbf{0.73}}} & \multicolumn{1}{c|}{\cellcolor[HTML]{FFABA7}{\color[HTML]{333333} \textbf{0.21}}} & \multicolumn{1}{c|}{\cellcolor[HTML]{D28986}{\color[HTML]{333333} \textbf{0.76}}} & \multicolumn{1}{c|}{\cellcolor[HTML]{D28986}{\color[HTML]{333333} \textbf{0.77}}} & \multicolumn{1}{c|}{\cellcolor[HTML]{D28986}{\color[HTML]{333333} \textbf{0.77}}} & \multicolumn{1}{c|}{\cellcolor[HTML]{FFFFFF}{\color[HTML]{333333} \textbf{0.08}}} & \multicolumn{1}{c|}{\cellcolor[HTML]{FFFFFF}{\color[HTML]{333333} \textbf{0.22}}} & \multicolumn{1}{c|}{\cellcolor[HTML]{FFFFFF}{\color[HTML]{333333} \textbf{0.24}}} & \multicolumn{1}{c|}{\cellcolor[HTML]{FFFFFF}{\color[HTML]{333333} \textbf{0.31}}} \\ \cline{3-12} 
\multicolumn{1}{|c|}{}                                                                                 & \multicolumn{1}{c|}{}                                                                 & \multicolumn{1}{c|}{OP}            & \multicolumn{1}{c|}{\cellcolor[HTML]{FFABA7}\textbf{0.72}}                        & \multicolumn{1}{c|}{\cellcolor[HTML]{FFABA7}\textbf{0.20}}                        & \multicolumn{1}{c|}{\cellcolor[HTML]{FFABA7}\textbf{0.74}}                        & \multicolumn{1}{c|}{\cellcolor[HTML]{D28986}\textbf{0.75}}                        & \multicolumn{1}{c|}{\cellcolor[HTML]{D28986}\textbf{0.75}}                        & \multicolumn{1}{c|}{\cellcolor[HTML]{D28986}\textbf{0.04}}                        & \multicolumn{1}{c|}{\cellcolor[HTML]{D28986}\textbf{0.05}}                        & \multicolumn{1}{c|}{\cellcolor[HTML]{D28986}\textbf{0.02}}                        & \multicolumn{1}{c|}{\cellcolor[HTML]{D28986}\textbf{0.04}}                        \\ \cline{3-12} 
\multicolumn{1}{|c|}{}                                                                                 & \multicolumn{1}{c|}{}                                                                 & \multicolumn{1}{c|}{Fairway}       & \multicolumn{1}{c|}{\cellcolor[HTML]{FFABA7}\textbf{0.71}}                        & \multicolumn{1}{c|}{\cellcolor[HTML]{D28986}\textbf{0.17}}                        & \multicolumn{1}{c|}{\cellcolor[HTML]{FFABA7}\textbf{0.73}}                        & \multicolumn{1}{c|}{\cellcolor[HTML]{FFABA7}\textbf{0.71}}                        & \multicolumn{1}{c|}{\cellcolor[HTML]{FFABA7}\textbf{0.71}}                        & \multicolumn{1}{c|}{\cellcolor[HTML]{D28986}\textbf{0.04}}                        & \multicolumn{1}{c|}{\cellcolor[HTML]{D28986}\textbf{0.03}}                        & \multicolumn{1}{c|}{\cellcolor[HTML]{D28986}\textbf{0.05}}                        & \multicolumn{1}{c|}{\cellcolor[HTML]{D28986}\textbf{0.06}}                        \\ \cline{3-12} 
\multicolumn{1}{|c|}{}                                                                                 & \multicolumn{1}{c|}{}                                                                 & \multicolumn{1}{c|}{Fair-SMOTE}    & \multicolumn{1}{c|}{\cellcolor[HTML]{D28986}{\color[HTML]{333333} \textbf{0.76}}} & \multicolumn{1}{c|}{\cellcolor[HTML]{D28986}{\color[HTML]{333333} \textbf{0.16}}} & \multicolumn{1}{c|}{\cellcolor[HTML]{FFABA7}{\color[HTML]{333333} \textbf{0.72}}} & \multicolumn{1}{c|}{\cellcolor[HTML]{FFABA7}{\color[HTML]{333333} \textbf{0.72}}} & \multicolumn{1}{c|}{\cellcolor[HTML]{D28986}{\color[HTML]{333333} \textbf{0.74}}} & \multicolumn{1}{c|}{\cellcolor[HTML]{D28986}{\color[HTML]{333333} \textbf{0.04}}} & \multicolumn{1}{c|}{\cellcolor[HTML]{D28986}{\color[HTML]{333333} \textbf{0.05}}} & \multicolumn{1}{c|}{\cellcolor[HTML]{D28986}{\color[HTML]{333333} \textbf{0.05}}} & \multicolumn{1}{c|}{\cellcolor[HTML]{D28986}{\color[HTML]{333333} \textbf{0.03}}} \\ \cline{3-12} 
\multicolumn{1}{|c|}{}                                                                                 & \multicolumn{1}{c|}{}                                                                 & \multicolumn{1}{c|}{Fair-SSL-ST}   & \multicolumn{1}{c|}{\cellcolor[HTML]{FFCCC9}{\color[HTML]{333333} \textbf{0.66}}} & \multicolumn{1}{c|}{\cellcolor[HTML]{FFFFFF}{\color[HTML]{333333} \textbf{0.37}}} & \multicolumn{1}{c|}{\cellcolor[HTML]{FFFFFF}{\color[HTML]{333333} \textbf{0.6}}}  & \multicolumn{1}{c|}{\cellcolor[HTML]{FFFFFF}{\color[HTML]{333333} \textbf{0.64}}} & \multicolumn{1}{c|}{\cellcolor[HTML]{FFCCC9}{\color[HTML]{333333} \textbf{0.64}}} & \multicolumn{1}{c|}{\cellcolor[HTML]{D28986}{\color[HTML]{333333} \textbf{0.04}}} & \multicolumn{1}{c|}{\cellcolor[HTML]{FFABA7}{\color[HTML]{333333} \textbf{0.08}}} & \multicolumn{1}{c|}{\cellcolor[HTML]{D28986}{\color[HTML]{333333} \textbf{0.05}}} & \multicolumn{1}{c|}{\cellcolor[HTML]{FFABA7}{\color[HTML]{333333} \textbf{0.11}}} \\ \cline{3-12} 
\multicolumn{1}{|c|}{}                                                                                 & \multicolumn{1}{c|}{}                                                                 & \multicolumn{1}{c|}{Fair-SSL-LP}   & \multicolumn{1}{c|}{\cellcolor[HTML]{FFFFFF}{\color[HTML]{333333} \textbf{0.57}}} & \multicolumn{1}{c|}{\cellcolor[HTML]{FFFFFF}{\color[HTML]{333333} \textbf{0.37}}} & \multicolumn{1}{c|}{\cellcolor[HTML]{FFFFFF}{\color[HTML]{333333} \textbf{0.58}}} & \multicolumn{1}{c|}{\cellcolor[HTML]{FFFFFF}{\color[HTML]{333333} \textbf{0.62}}} & \multicolumn{1}{c|}{\cellcolor[HTML]{FFFFFF}{\color[HTML]{333333} \textbf{0.56}}} & \multicolumn{1}{c|}{\cellcolor[HTML]{D28986}{\color[HTML]{333333} \textbf{0.02}}} & \multicolumn{1}{c|}{\cellcolor[HTML]{D28986}{\color[HTML]{333333} \textbf{0.03}}} & \multicolumn{1}{c|}{\cellcolor[HTML]{D28986}{\color[HTML]{333333} \textbf{0.07}}} & \multicolumn{1}{c|}{\cellcolor[HTML]{FFABA7}{\color[HTML]{333333} \textbf{0.1}}}  \\ \cline{3-12} 
\multicolumn{1}{|c|}{}                                                                                 & \multicolumn{1}{c|}{}                                                                 & \multicolumn{1}{c|}{Fair-SSL-LS}   & \multicolumn{1}{c|}{\cellcolor[HTML]{FFFFFF}{\color[HTML]{333333} \textbf{0.62}}} & \multicolumn{1}{c|}{\cellcolor[HTML]{FFFFFF}{\color[HTML]{333333} \textbf{0.35}}} & \multicolumn{1}{c|}{\cellcolor[HTML]{FFFFFF}{\color[HTML]{333333} \textbf{0.62}}} & \multicolumn{1}{c|}{\cellcolor[HTML]{FFFFFF}{\color[HTML]{333333} \textbf{0.64}}} & \multicolumn{1}{c|}{\cellcolor[HTML]{FFCCC9}{\color[HTML]{333333} \textbf{0.62}}} & \multicolumn{1}{c|}{\cellcolor[HTML]{D28986}{\color[HTML]{333333} \textbf{0.07}}} & \multicolumn{1}{c|}{\cellcolor[HTML]{D28986}{\color[HTML]{333333} \textbf{0.09}}} & \multicolumn{1}{c|}{\cellcolor[HTML]{D28986}{\color[HTML]{333333} \textbf{0.07}}} & \multicolumn{1}{c|}{\cellcolor[HTML]{FFABA7}{\color[HTML]{333333} \textbf{0.12}}} \\ \cline{3-12} 
\multicolumn{1}{|c|}{\multirow{-8}{*}{\begin{tabular}[c]{@{}c@{}}Bank \\ Marketing\end{tabular}}}      & \multicolumn{1}{c|}{\multirow{-8}{*}{Age}}                                            & \multicolumn{1}{c|}{Fair-SSL-CT}   & \multicolumn{1}{c|}{\cellcolor[HTML]{FFCCC9}{\color[HTML]{333333} \textbf{0.69}}} & \multicolumn{1}{c|}{\cellcolor[HTML]{FFABA7}{\color[HTML]{333333} \textbf{0.18}}} & \multicolumn{1}{c|}{\cellcolor[HTML]{D28986}{\color[HTML]{333333} \textbf{0.74}}} & \multicolumn{1}{c|}{\cellcolor[HTML]{FFABA7}{\color[HTML]{333333} \textbf{0.72}}} & \multicolumn{1}{c|}{\cellcolor[HTML]{FFABA7}{\color[HTML]{333333} \textbf{0.71}}} & \multicolumn{1}{c|}{\cellcolor[HTML]{D28986}{\color[HTML]{333333} \textbf{0.05}}} & \multicolumn{1}{c|}{\cellcolor[HTML]{D28986}{\color[HTML]{333333} \textbf{0.02}}} & \multicolumn{1}{c|}{\cellcolor[HTML]{D28986}{\color[HTML]{333333} \textbf{0.09}}} & \multicolumn{1}{c|}{\cellcolor[HTML]{FFCCC9}{\color[HTML]{333333} \textbf{0.21}}} \\ \hline
\multicolumn{1}{|c|}{}                                                                                 & \multicolumn{1}{c|}{}                                                                 & \multicolumn{1}{c|}{Default}       & \multicolumn{1}{c|}{\cellcolor[HTML]{FFCCC9}{\color[HTML]{333333} \textbf{0.81}}} & \multicolumn{1}{c|}{\cellcolor[HTML]{D28986}{\color[HTML]{333333} \textbf{0.06}}} & \multicolumn{1}{c|}{\cellcolor[HTML]{D28986}{\color[HTML]{333333} \textbf{0.85}}} & \multicolumn{1}{c|}{\cellcolor[HTML]{D28986}{\color[HTML]{333333} \textbf{0.88}}} & \multicolumn{1}{c|}{\cellcolor[HTML]{FFFFFF}{\color[HTML]{333333} \textbf{0.83}}} & \multicolumn{1}{c|}{\cellcolor[HTML]{FFABA7}{\color[HTML]{333333} \textbf{0.06}}} & \multicolumn{1}{c|}{\cellcolor[HTML]{FFFFFF}{\color[HTML]{333333} \textbf{0.05}}} & \multicolumn{1}{c|}{\cellcolor[HTML]{FFFFFF}{\color[HTML]{333333} \textbf{0.06}}} & \multicolumn{1}{c|}{\cellcolor[HTML]{FFFFFF}{\color[HTML]{333333} \textbf{0.12}}} \\ \cline{3-12} 
\multicolumn{1}{|c|}{}                                                                                 & \multicolumn{1}{c|}{}                                                                 & \multicolumn{1}{c|}{OP}            & \multicolumn{1}{c|}{\cellcolor[HTML]{FFCCC9}{\color[HTML]{333333} \textbf{0.79}}} & \multicolumn{1}{c|}{\cellcolor[HTML]{D28986}\textbf{0.06}}                        & \multicolumn{1}{c|}{\cellcolor[HTML]{D28986}\textbf{0.83}}                        & \multicolumn{1}{c|}{\cellcolor[HTML]{FFABA7}\textbf{0.83}}                        & \multicolumn{1}{c|}{\textbf{0.82}}                                                & \multicolumn{1}{c|}{\cellcolor[HTML]{D28986}\textbf{0.02}}                        & \multicolumn{1}{c|}{\cellcolor[HTML]{D28986}\textbf{0.04}}                        & \multicolumn{1}{c|}{\cellcolor[HTML]{D28986}\textbf{0.04}}                        & \multicolumn{1}{c|}{\cellcolor[HTML]{D28986}\textbf{0.06}}                        \\ \cline{3-12} 
\multicolumn{1}{|c|}{}                                                                                 & \multicolumn{1}{c|}{}                                                                 & \multicolumn{1}{c|}{Fairway}       & \multicolumn{1}{c|}{\textbf{0.76}}                                                & \multicolumn{1}{c|}{\cellcolor[HTML]{D28986}\textbf{0.05}}                        & \multicolumn{1}{c|}{\cellcolor[HTML]{FFABA7}\textbf{0.81}}                        & \multicolumn{1}{c|}{\cellcolor[HTML]{FFABA7}\textbf{0.84}}                        & \multicolumn{1}{c|}{\textbf{0.84}}                                                & \multicolumn{1}{c|}{\cellcolor[HTML]{D28986}\textbf{0.03}}                        & \multicolumn{1}{c|}{\cellcolor[HTML]{D28986}\textbf{0.02}}                        & \multicolumn{1}{c|}{\cellcolor[HTML]{D28986}\textbf{0.04}}                        & \multicolumn{1}{c|}{\cellcolor[HTML]{D28986}\textbf{0.07}}                        \\ \cline{3-12} 
\multicolumn{1}{|c|}{}                                                                                 & \multicolumn{1}{c|}{}                                                                 & \multicolumn{1}{c|}{Fair-SMOTE}    & \multicolumn{1}{c|}{\cellcolor[HTML]{D28986}{\color[HTML]{333333} \textbf{0.87}}} & \multicolumn{1}{c|}{\cellcolor[HTML]{FFABA7}{\color[HTML]{333333} \textbf{0.10}}} & \multicolumn{1}{c|}{\cellcolor[HTML]{D28986}{\color[HTML]{333333} \textbf{0.84}}} & \multicolumn{1}{c|}{\cellcolor[HTML]{D28986}{\color[HTML]{333333} \textbf{0.87}}} & \multicolumn{1}{c|}{\cellcolor[HTML]{D28986}{\color[HTML]{333333} \textbf{0.86}}} & \multicolumn{1}{c|}{\cellcolor[HTML]{D28986}{\color[HTML]{333333} \textbf{0.04}}} & \multicolumn{1}{c|}{\cellcolor[HTML]{D28986}{\color[HTML]{333333} \textbf{0.04}}} & \multicolumn{1}{c|}{\cellcolor[HTML]{D28986}{\color[HTML]{333333} \textbf{0.04}}} & \multicolumn{1}{c|}{\cellcolor[HTML]{D28986}{\color[HTML]{333333} \textbf{0.08}}} \\ \cline{3-12} 
\multicolumn{1}{|c|}{}                                                                                 & \multicolumn{1}{c|}{}                                                                 & \multicolumn{1}{c|}{Fair-SSL-ST}   & \multicolumn{1}{c|}{\cellcolor[HTML]{FFABA7}{\color[HTML]{333333} \textbf{0.84}}} & \multicolumn{1}{c|}{\cellcolor[HTML]{FFFFFF}{\color[HTML]{333333} \textbf{0.14}}} & \multicolumn{1}{c|}{\cellcolor[HTML]{D28986}{\color[HTML]{333333} \textbf{0.83}}} & \multicolumn{1}{c|}{\cellcolor[HTML]{FFABA7}{\color[HTML]{333333} \textbf{0.83}}} & \multicolumn{1}{c|}{\cellcolor[HTML]{FFFFFF}{\color[HTML]{333333} \textbf{0.83}}} & \multicolumn{1}{c|}{\cellcolor[HTML]{D28986}{\color[HTML]{333333} \textbf{0.03}}} & \multicolumn{1}{c|}{\cellcolor[HTML]{D28986}{\color[HTML]{333333} \textbf{0.02}}} & \multicolumn{1}{c|}{\cellcolor[HTML]{D28986}{\color[HTML]{333333} \textbf{0.01}}} & \multicolumn{1}{c|}{\cellcolor[HTML]{D28986}{\color[HTML]{333333} \textbf{0.03}}} \\ \cline{3-12} 
\multicolumn{1}{|c|}{}                                                                                 & \multicolumn{1}{c|}{}                                                                 & \multicolumn{1}{c|}{Fair-SSL-LP}   & \multicolumn{1}{c|}{\cellcolor[HTML]{FFABA7}{\color[HTML]{333333} \textbf{0.83}}} & \multicolumn{1}{c|}{\cellcolor[HTML]{FFFFFF}{\color[HTML]{333333} \textbf{0.19}}} & \multicolumn{1}{c|}{\cellcolor[HTML]{D28986}{\color[HTML]{333333} \textbf{0.82}}} & \multicolumn{1}{c|}{\cellcolor[HTML]{FFFFFF}{\color[HTML]{333333} \textbf{0.82}}} & \multicolumn{1}{c|}{\cellcolor[HTML]{FFFFFF}{\color[HTML]{333333} \textbf{0.83}}} & \multicolumn{1}{c|}{\cellcolor[HTML]{D28986}{\color[HTML]{333333} \textbf{0.04}}} & \multicolumn{1}{c|}{\cellcolor[HTML]{D28986}{\color[HTML]{333333} \textbf{0.03}}} & \multicolumn{1}{c|}{\cellcolor[HTML]{D28986}{\color[HTML]{333333} \textbf{0.07}}} & \multicolumn{1}{c|}{\cellcolor[HTML]{FFABA7}{\color[HTML]{333333} \textbf{0.21}}} \\ \cline{3-12} 
\multicolumn{1}{|c|}{}                                                                                 & \multicolumn{1}{c|}{}                                                                 & \multicolumn{1}{c|}{Fair-SSL-LS}   & \multicolumn{1}{c|}{\cellcolor[HTML]{FFABA7}\textbf{0.84}}                        & \multicolumn{1}{c|}{\textbf{0.14}}                                                & \multicolumn{1}{c|}{\textbf{0.79}}                                                & \multicolumn{1}{c|}{\cellcolor[HTML]{FFABA7}\textbf{0.84}}                        & \multicolumn{1}{c|}{\textbf{0.82}}                                                & \multicolumn{1}{c|}{\cellcolor[HTML]{D28986}\textbf{0.03}}                        & \multicolumn{1}{c|}{\cellcolor[HTML]{D28986}\textbf{0.02}}                        & \multicolumn{1}{c|}{\cellcolor[HTML]{D28986}\textbf{0.05}}                        & \multicolumn{1}{c|}{\cellcolor[HTML]{D28986}\textbf{0.05}}                        \\ \cline{3-12} 
\multicolumn{1}{|c|}{\multirow{-8}{*}{\begin{tabular}[c]{@{}c@{}}Student\\ Performance\end{tabular}}}  & \multicolumn{1}{c|}{\multirow{-8}{*}{Sex}}                                            & \multicolumn{1}{c|}{Fair-SSL-CT}   & \multicolumn{1}{c|}{\cellcolor[HTML]{FFABA7}\textbf{0.84}}                        & \multicolumn{1}{c|}{\cellcolor[HTML]{FFABA7}\textbf{0.09}}                        & \multicolumn{1}{c|}{\cellcolor[HTML]{D28986}\textbf{0.84}}                        & \multicolumn{1}{c|}{\cellcolor[HTML]{FFABA7}\textbf{0.85}}                        & \multicolumn{1}{c|}{\textbf{0.83}}                                                & \multicolumn{1}{c|}{\cellcolor[HTML]{D28986}\textbf{0.04}}                        & \multicolumn{1}{c|}{\cellcolor[HTML]{D28986}\textbf{0.02}}                        & \multicolumn{1}{c|}{\cellcolor[HTML]{D28986}\textbf{0.03}}                        & \multicolumn{1}{c|}{\cellcolor[HTML]{D28986}\textbf{0.08}}                        \\ \hline
\end{tabular}
\end{table*}

\begin{table*}[!ht]
\scriptsize
\caption{RQ2 results: Comparing Fair-SSL with three prior works based on results of 10 datasets and three learners. Number of wins, ties, and losses are calculated using Scott-Knott ranks. \colorbox{pink}{Pink} cells show Fair-SSL is performing similar/better than others.}
\label{Algo_comparison}
\begin{tabular}{
>{\columncolor[HTML]{C0C0C0}}c 
>{\columncolor[HTML]{C0C0C0}}c cccccccccc}
\cline{3-12}
\multicolumn{1}{l}{\cellcolor[HTML]{C0C0C0}} &                                                                 & \cellcolor[HTML]{C0C0C0}\textbf{Recall}         & \cellcolor[HTML]{C0C0C0}\textbf{False alarm} & \cellcolor[HTML]{C0C0C0}\textbf{Precision} & \cellcolor[HTML]{C0C0C0}\textbf{Accuracy} & \cellcolor[HTML]{C0C0C0}\textbf{F1 Score}       & \cellcolor[HTML]{C0C0C0}\textbf{AOD}            & \cellcolor[HTML]{C0C0C0}\textbf{EOD}            & \cellcolor[HTML]{C0C0C0}\textbf{SPD}            & \cellcolor[HTML]{C0C0C0}\textbf{DI}             & \cellcolor[HTML]{C0C0C0}\textbf{Total} \\ \cline{3-12} 
\multicolumn{1}{l}{\cellcolor[HTML]{C0C0C0}} & \multicolumn{11}{c}{\cellcolor[HTML]{C0C0C0}\textbf{Optimized Pre-processing vs Fair-SSL}}                                                                                                                                                                                                                                                                                                                                                                                                                                                                   \\ \cline{3-12} 
1                                            & \multicolumn{1}{c|}{\cellcolor[HTML]{C0C0C0}\textbf{Win}}       & \multicolumn{1}{c|}{\cellcolor[HTML]{FFABA7}8}  & \multicolumn{1}{c|}{2}                       & \multicolumn{1}{c|}{4}                     & \multicolumn{1}{c|}{3}                    & \multicolumn{1}{c|}{\cellcolor[HTML]{FFABA7}10} & \multicolumn{1}{c|}{\cellcolor[HTML]{FFABA7}2}  & \multicolumn{1}{c|}{\cellcolor[HTML]{FFABA7}2}  & \multicolumn{1}{c|}{\cellcolor[HTML]{FFABA7}1}  & \multicolumn{1}{c|}{\cellcolor[HTML]{FFABA7}2}  & \multicolumn{1}{c|}{34}                \\ \cline{3-12} 
2                                            & \multicolumn{1}{c|}{\cellcolor[HTML]{C0C0C0}\textbf{Tie}}       & \multicolumn{1}{c|}{\cellcolor[HTML]{FFABA7}21} & \multicolumn{1}{c|}{25}                      & \multicolumn{1}{c|}{27}                    & \multicolumn{1}{c|}{28}                   & \multicolumn{1}{c|}{\cellcolor[HTML]{FFABA7}24} & \multicolumn{1}{c|}{\cellcolor[HTML]{FFABA7}32} & \multicolumn{1}{c|}{\cellcolor[HTML]{FFABA7}33} & \multicolumn{1}{c|}{\cellcolor[HTML]{FFABA7}32} & \multicolumn{1}{c|}{\cellcolor[HTML]{FFABA7}32} & \multicolumn{1}{c|}{254}               \\ \cline{3-12} 
3                                            & \multicolumn{1}{c|}{\cellcolor[HTML]{C0C0C0}\textbf{Loss}}      & \multicolumn{1}{c|}{7}                          & \multicolumn{1}{c|}{9}                       & \multicolumn{1}{c|}{5}                     & \multicolumn{1}{c|}{5}                    & \multicolumn{1}{c|}{2}                          & \multicolumn{1}{c|}{2}                          & \multicolumn{1}{c|}{1}                          & \multicolumn{1}{c|}{3}                          & \multicolumn{1}{c|}{2}                          & \multicolumn{1}{c|}{36}                \\ \cline{3-12} 
4                                            & \multicolumn{1}{c|}{\cellcolor[HTML]{C0C0C0}\textbf{Win + Tie}} & \multicolumn{1}{c|}{29}                         & \multicolumn{1}{c|}{27}                      & \multicolumn{1}{c|}{31}                    & \multicolumn{1}{c|}{31}                   & \multicolumn{1}{c|}{34}                         & \multicolumn{1}{c|}{\cellcolor[HTML]{FFABA7}34} & \multicolumn{1}{c|}{\cellcolor[HTML]{FFABA7}35} & \multicolumn{1}{c|}{\cellcolor[HTML]{FFABA7}33} & \multicolumn{1}{c|}{\cellcolor[HTML]{FFABA7}34} & \multicolumn{1}{c|}{288/324}           \\ \cline{3-12} 
                                             & \multicolumn{11}{c}{\cellcolor[HTML]{C0C0C0}\textbf{Fairway vs Fair-SSL}}                                                                                                                                                                                                                                                                                                                                                                                                                                                                                    \\ \cline{3-12} 
5                                            & \multicolumn{1}{c|}{\cellcolor[HTML]{C0C0C0}\textbf{Win}}       & \multicolumn{1}{c|}{\cellcolor[HTML]{FFFFFF}10} & \multicolumn{1}{c|}{2}                       & \multicolumn{1}{c|}{5}                     & \multicolumn{1}{c|}{5}                    & \multicolumn{1}{c|}{\cellcolor[HTML]{FFABA7}19} & \multicolumn{1}{c|}{\cellcolor[HTML]{FFABA7}2}  & \multicolumn{1}{c|}{\cellcolor[HTML]{FFABA7}3}  & \multicolumn{1}{c|}{\cellcolor[HTML]{FFABA7}3}  & \multicolumn{1}{c|}{\cellcolor[HTML]{FFABA7}4}  & \multicolumn{1}{c|}{53}                \\ \cline{3-12} 
6                                            & \multicolumn{1}{c|}{\cellcolor[HTML]{C0C0C0}\textbf{Tie}}       & \multicolumn{1}{c|}{\cellcolor[HTML]{FFABA7}24} & \multicolumn{1}{c|}{17}                      & \multicolumn{1}{c|}{24}                    & \multicolumn{1}{c|}{27}                   & \multicolumn{1}{c|}{\cellcolor[HTML]{FFABA7}14} & \multicolumn{1}{c|}{\cellcolor[HTML]{FFABA7}30} & \multicolumn{1}{c|}{\cellcolor[HTML]{FFABA7}31} & \multicolumn{1}{c|}{\cellcolor[HTML]{FFABA7}32} & \multicolumn{1}{c|}{\cellcolor[HTML]{FFABA7}31} & \multicolumn{1}{c|}{230}               \\ \cline{3-12} 
7                                            & \multicolumn{1}{c|}{\cellcolor[HTML]{C0C0C0}\textbf{Loss}}      & \multicolumn{1}{c|}{2}                          & \multicolumn{1}{c|}{17}                      & \multicolumn{1}{c|}{7}                     & \multicolumn{1}{c|}{4}                    & \multicolumn{1}{c|}{3}                          & \multicolumn{1}{c|}{4}                          & \multicolumn{1}{c|}{2}                          & \multicolumn{1}{c|}{1}                          & \multicolumn{1}{c|}{1}                          & \multicolumn{1}{c|}{41}                \\ \cline{3-12} 
8                                            & \multicolumn{1}{c|}{\cellcolor[HTML]{C0C0C0}\textbf{Win + Tie}} & \multicolumn{1}{c|}{\cellcolor[HTML]{FFABA7}34} & \multicolumn{1}{c|}{19}                      & \multicolumn{1}{c|}{29}                    & \multicolumn{1}{c|}{33}                   & \multicolumn{1}{c|}{\cellcolor[HTML]{FFABA7}33} & \multicolumn{1}{c|}{\cellcolor[HTML]{FFABA7}32} & \multicolumn{1}{c|}{\cellcolor[HTML]{FFABA7}34} & \multicolumn{1}{c|}{\cellcolor[HTML]{FFABA7}35} & \multicolumn{1}{c|}{\cellcolor[HTML]{FFABA7}35} & \multicolumn{1}{c|}{283/324}           \\ \cline{3-12} 
                                             & \multicolumn{11}{c}{\cellcolor[HTML]{C0C0C0}\textbf{Fair-SMOTE vs Fair-SSL}}                                                                                                                                                                                                                                                                                                                                                                                                                                                                                 \\ \cline{3-12} 
9                                            & \multicolumn{1}{c|}{\cellcolor[HTML]{C0C0C0}\textbf{Win}}       & \multicolumn{1}{c|}{\cellcolor[HTML]{FFFFFF}1}  & \multicolumn{1}{c|}{3}                       & \multicolumn{1}{c|}{2}                     & \multicolumn{1}{c|}{3}                    & \multicolumn{1}{c|}{\cellcolor[HTML]{FFFFFF}3}  & \multicolumn{1}{c|}{\cellcolor[HTML]{FFABA7}1}  & \multicolumn{1}{c|}{\cellcolor[HTML]{FFABA7}2}  & \multicolumn{1}{c|}{\cellcolor[HTML]{FFABA7}2}  & \multicolumn{1}{c|}{\cellcolor[HTML]{FFABA7}2}  & \multicolumn{1}{c|}{19}                \\ \cline{3-12} 
10                                           & \multicolumn{1}{c|}{\cellcolor[HTML]{C0C0C0}\textbf{Tie}}       & \multicolumn{1}{c|}{\cellcolor[HTML]{FFFFFF}28} & \multicolumn{1}{c|}{30}                      & \multicolumn{1}{c|}{30}                    & \multicolumn{1}{c|}{28}                   & \multicolumn{1}{c|}{\cellcolor[HTML]{FFFFFF}29} & \multicolumn{1}{c|}{\cellcolor[HTML]{FFABA7}34} & \multicolumn{1}{c|}{\cellcolor[HTML]{FFABA7}33} & \multicolumn{1}{c|}{\cellcolor[HTML]{FFABA7}32} & \multicolumn{1}{c|}{\cellcolor[HTML]{FFABA7}31} & \multicolumn{1}{c|}{275}               \\ \cline{3-12} 
11                                           & \multicolumn{1}{c|}{\cellcolor[HTML]{C0C0C0}\textbf{Loss}}      & \multicolumn{1}{c|}{7}                          & \multicolumn{1}{c|}{3}                       & \multicolumn{1}{c|}{4}                     & \multicolumn{1}{c|}{5}                    & \multicolumn{1}{c|}{4}                          & \multicolumn{1}{c|}{1}                          & \multicolumn{1}{c|}{1}                          & \multicolumn{1}{c|}{2}                          & \multicolumn{1}{c|}{3}                          & \multicolumn{1}{c|}{30}                \\ \cline{3-12} 
12                                           & \multicolumn{1}{c|}{\cellcolor[HTML]{C0C0C0}\textbf{Win + Tie}} & \multicolumn{1}{c|}{\cellcolor[HTML]{FFFFFF}29} & \multicolumn{1}{c|}{33}                      & \multicolumn{1}{c|}{32}                    & \multicolumn{1}{c|}{31}                   & \multicolumn{1}{c|}{\cellcolor[HTML]{FFFFFF}32} & \multicolumn{1}{c|}{\cellcolor[HTML]{FFABA7}35} & \multicolumn{1}{c|}{\cellcolor[HTML]{FFABA7}35} & \multicolumn{1}{c|}{\cellcolor[HTML]{FFABA7}34} & \multicolumn{1}{c|}{\cellcolor[HTML]{FFABA7}33} & \multicolumn{1}{c|}{294/324}           \\ \cline{3-12} 
\end{tabular}
\end{table*}

% \vspace{-1cm}

\begin{table}[]
\caption{RQ3 results: The change of accuracy, F1, AOD and EOD for Fair-SSL with increasing size of labeled training data.\colorbox{deeppink}{Dark} = 1st Rank;\colorbox{pink}{Light} = 2nd Rank; White = Last Rank}
\label{Initial_size}
\scriptsize
\adjustbox{width=3.3in}{
\begin{tabular}{ccccc
>{\columncolor[HTML]{D28986}}c c}
\cellcolor[HTML]{C0C0C0}Dataset                 & \cellcolor[HTML]{C0C0C0}\begin{tabular}[c]{@{}c@{}}Protected\\ Attribute\end{tabular} & \cellcolor[HTML]{C0C0C0}\begin{tabular}[c]{@{}c@{}}Size of\\ Labeled\\ Set\end{tabular} & \cellcolor[HTML]{C0C0C0}\begin{tabular}[c]{@{}c@{}}Accuracy\\ (+)\end{tabular}    & \cellcolor[HTML]{C0C0C0}\begin{tabular}[c]{@{}c@{}}F1 Score\\ (+)\end{tabular}    & \cellcolor[HTML]{C0C0C0}\begin{tabular}[c]{@{}c@{}}AOD\\ (-)\end{tabular}         & \cellcolor[HTML]{C0C0C0}\begin{tabular}[c]{@{}c@{}}EOD\\ (-)\end{tabular}         \\ \hline
\multicolumn{1}{|c|}{}                          & \multicolumn{1}{c|}{}                                                                 & \multicolumn{1}{c|}{1\%}                                                                & \multicolumn{1}{c|}{\cellcolor[HTML]{FFFFFF}{\color[HTML]{333333} \textbf{0.61}}} & \multicolumn{1}{c|}{\cellcolor[HTML]{FFFFFF}{\color[HTML]{333333} \textbf{0.44}}} & \multicolumn{1}{c|}{\cellcolor[HTML]{FFFFFF}{\color[HTML]{333333} \textbf{0.04}}} & \multicolumn{1}{c|}{\cellcolor[HTML]{FFFFFF}{\color[HTML]{333333} \textbf{0.06}}} \\ \cline{3-7} 
\multicolumn{1}{|c|}{}                          & \multicolumn{1}{c|}{}                                                                 & \multicolumn{1}{c|}{5\%}                                                                & \multicolumn{1}{c|}{\cellcolor[HTML]{FFABA7}\textbf{0.68}}                        & \multicolumn{1}{c|}{\cellcolor[HTML]{FFFFFF}\textbf{0.51}}                        & \multicolumn{1}{c|}{\cellcolor[HTML]{D28986}\textbf{0.03}}                        & \multicolumn{1}{c|}{\cellcolor[HTML]{D28986}\textbf{0.02}}                        \\ \cline{3-7} 
\multicolumn{1}{|c|}{}                          & \multicolumn{1}{c|}{}                                                                 & \multicolumn{1}{c|}{10\%}                                                               & \multicolumn{1}{c|}{\cellcolor[HTML]{D28986}\textbf{0.71}}                        & \multicolumn{1}{c|}{\cellcolor[HTML]{FFABA7}\textbf{0.54}}                        & \multicolumn{1}{c|}{\cellcolor[HTML]{D28986}\textbf{0.03}}                        & \multicolumn{1}{c|}{\cellcolor[HTML]{D28986}\textbf{0.03}}                        \\ \cline{3-7} 
\multicolumn{1}{|c|}{\multirow{-4}{*}{Adult}}   & \multicolumn{1}{c|}{\multirow{-4}{*}{Sex}}                                            & \multicolumn{1}{c|}{20\%}                                                               & \multicolumn{1}{c|}{\cellcolor[HTML]{D28986}{\color[HTML]{333333} \textbf{0.71}}} & \multicolumn{1}{c|}{\cellcolor[HTML]{D28986}{\color[HTML]{333333} \textbf{0.58}}} & \multicolumn{1}{c|}{\cellcolor[HTML]{D28986}{\color[HTML]{333333} \textbf{0.02}}} & \multicolumn{1}{c|}{\cellcolor[HTML]{D28986}{\color[HTML]{333333} \textbf{0.02}}} \\ \hline
\multicolumn{1}{|c|}{}                          & \multicolumn{1}{c|}{}                                                                 & \multicolumn{1}{c|}{1\%}                                                                & \multicolumn{1}{c|}{\cellcolor[HTML]{FFFFFF}\textbf{0.47}}                        & \multicolumn{1}{c|}{\cellcolor[HTML]{FFFFFF}\textbf{0.44}}                        & \multicolumn{1}{c|}{\cellcolor[HTML]{FFFFFF}\textbf{0.07}}                        & \multicolumn{1}{c|}{\cellcolor[HTML]{FFFFFF}\textbf{0.12}}                        \\ \cline{3-7} 
\multicolumn{1}{|c|}{}                          & \multicolumn{1}{c|}{}                                                                 & \multicolumn{1}{c|}{5\%}                                                                & \multicolumn{1}{c|}{\cellcolor[HTML]{FFABA7}\textbf{0.57}}                        & \multicolumn{1}{c|}{\cellcolor[HTML]{FFFFFF}\textbf{0.54}}                        & \multicolumn{1}{c|}{\cellcolor[HTML]{FFFFFF}\textbf{0.06}}                        & \multicolumn{1}{c|}{\cellcolor[HTML]{FFABA7}\textbf{0.06}}                        \\ \cline{3-7} 
\multicolumn{1}{|c|}{}                          & \multicolumn{1}{c|}{}                                                                 & \multicolumn{1}{c|}{10\%}                                                               & \multicolumn{1}{c|}{\cellcolor[HTML]{FFABA7}{\color[HTML]{333333} \textbf{0.58}}} & \multicolumn{1}{c|}{\cellcolor[HTML]{FFABA7}{\color[HTML]{333333} \textbf{0.61}}} & \multicolumn{1}{c|}{\cellcolor[HTML]{D28986}{\color[HTML]{333333} \textbf{0.03}}} & \multicolumn{1}{c|}{\cellcolor[HTML]{FFABA7}{\color[HTML]{333333} \textbf{0.07}}} \\ \cline{3-7} 
\multicolumn{1}{|c|}{\multirow{-4}{*}{Compas}}  & \multicolumn{1}{c|}{\multirow{-4}{*}{Sex}}                                            & \multicolumn{1}{c|}{20\%}                                                               & \multicolumn{1}{c|}{\cellcolor[HTML]{D28986}{\color[HTML]{333333} \textbf{0.61}}} & \multicolumn{1}{c|}{\cellcolor[HTML]{D28986}{\color[HTML]{333333} \textbf{0.64}}} & \multicolumn{1}{c|}{\cellcolor[HTML]{D28986}{\color[HTML]{333333} \textbf{0.01}}} & \multicolumn{1}{c|}{\cellcolor[HTML]{D28986}{\color[HTML]{333333} \textbf{0.05}}} \\ \hline
\multicolumn{1}{|c|}{}                          & \multicolumn{1}{c|}{}                                                                 & \multicolumn{1}{c|}{1\%}                                                                & \multicolumn{1}{c|}{\cellcolor[HTML]{FFFFFF}\textbf{0.67}}                        & \multicolumn{1}{c|}{\cellcolor[HTML]{FFFFFF}\textbf{0.65}}                        & \multicolumn{1}{c|}{\cellcolor[HTML]{FFFFFF}\textbf{0.18}}                        & \multicolumn{1}{c|}{\cellcolor[HTML]{FFFFFF}\textbf{0.12}}                        \\ \cline{3-7} 
\multicolumn{1}{|c|}{}                          & \multicolumn{1}{c|}{}                                                                 & \multicolumn{1}{c|}{5\%}                                                                & \multicolumn{1}{c|}{\cellcolor[HTML]{FFABA7}\textbf{0.78}}                        & \multicolumn{1}{c|}{\cellcolor[HTML]{FFABA7}\textbf{0.75}}                        & \multicolumn{1}{c|}{\cellcolor[HTML]{FFABA7}\textbf{0.05}}                        & \multicolumn{1}{c|}{\cellcolor[HTML]{FFABA7}\textbf{0.07}}                        \\ \cline{3-7} 
\multicolumn{1}{|c|}{}                          & \multicolumn{1}{c|}{}                                                                 & \multicolumn{1}{c|}{10\%}                                                               & \multicolumn{1}{c|}{\cellcolor[HTML]{D28986}{\color[HTML]{333333} \textbf{0.83}}} & \multicolumn{1}{c|}{\cellcolor[HTML]{D28986}{\color[HTML]{333333} \textbf{0.82}}} & \multicolumn{1}{c|}{\cellcolor[HTML]{D28986}{\color[HTML]{333333} \textbf{0.02}}} & \multicolumn{1}{c|}{\cellcolor[HTML]{D28986}{\color[HTML]{333333} \textbf{0.02}}} \\ \cline{3-7} 
\multicolumn{1}{|c|}{\multirow{-4}{*}{Student}} & \multicolumn{1}{c|}{\multirow{-4}{*}{Sex}}                                            & \multicolumn{1}{c|}{20\%}                                                               & \multicolumn{1}{c|}{\cellcolor[HTML]{D28986}{\color[HTML]{333333} \textbf{0.85}}} & \multicolumn{1}{c|}{\cellcolor[HTML]{D28986}{\color[HTML]{333333} \textbf{0.83}}} & \multicolumn{1}{c|}{\cellcolor[HTML]{D28986}{\color[HTML]{333333} \textbf{0.02}}} & \multicolumn{1}{c|}{\cellcolor[HTML]{D28986}{\color[HTML]{333333} \textbf{0.01}}} \\ \hline
\end{tabular}}
\end{table}
\vspace{-.25cm}

\newenvironment{RQ}{\vspace{2mm}\begin{tcolorbox}[enhanced,width=3.3in,size=fbox,colback=blue!5,drop shadow southeast,sharp corners]}{\end{tcolorbox}}

\begin{RQ}
{\bf RQ1.} Can Fair-SSL reduce bias? 
\end{RQ}
Table \ref{RQ2_3} answers the question. It contains the results for five datasets. ``Default'' rows signify when a logistic regression model is trained on the available labeled training data with no modification. We see for every dataset, the values of AOD, EOD, SPD, and DI are very high indicating bias in prediction. For every dataset, the last four rows show the results of Fair-SSL with four different algorithms used as pseudo-labeler inside. That means, 10\% labeled training data is used and rest of the training data has been pseudo-labeled. After that combined training data is oversampled to be balanced based on protected attribute and class. Finally, the generated data has been used for model training. We see all four versions of Fair-SSL significantly reduce the values of four bias metrics (AOD, EOD, SPD, and DI) for all the datasets. The answer  for RQ1 is \textbf{``Yes, Fair-SSL can significantly reduce bias. It improves recall and F1 and sometimes damages false alarm, and accuracy. Prior fairness works~\cite{Chakraborty_2020,Kamiran:2018:ERO:3165328.3165686,Kamiran2012,zhang2018mitigating,NIPS2017_6988} also damaged performance of the model while achieving fairness. This is called the ``fairness-performance'' trade-off in the literature.''}

\begin{RQ}
{\bf RQ2.} How well does Fair-SSL perform compared to the state of the art bias mitigation algorithms? 
\end{RQ}
Here we are comparing Fair-SSL with three other state-of-the-art bias mitigation algorithms - Optimized Pre-processing~\cite{NIPS2017_6988}, Fairway~\cite{Chakraborty_2020} and Fair-SMOTE~\cite{Chakraborty_2020}. If we look at Table \ref{RQ2_3}, the last four columns are bias scores where Fair-SSL is performing as good as the others. In case of performance metrics (first five columns), we see it is losing sometimes but not by much. To get a clear picture, we take a look at Table \ref{Algo_comparison}. Here we show the summarized results for all ten datasets (Adult and Compas have two protected attributes each, i.e. 10 + 2 = 12 cases) and three learners (logistic regression, random forest, and svm). We have implemented four different versions of Fair-SSL (ST, LP, LS, CT). For comparison purpose, for every dataset, we have created a validation set (20\%) from the train set. On the validation set, we tried all four versions of Fair-SSL and chose the best one to run on the test set. Thus, Table \ref{Algo_comparison} contains comparison of the best version of Fair-SSL (best is selected based on validation results)  with three other bias mitigation algorithms. We see in bias scores, Fair-SSL is as good as other three. In `recall' and `F1 score', it performs better than Optimized Pre-processing and Fairway. One important point to mention here is that other three algorithms take advantage of full training data where Fair-SSL takes only 10\% of labeled training data. The answer for RQ2 is \textbf{``Fair-SSL is as good as others in reducing bias, sometimes better in ``recall'' and ``F1'' than Optimized Pre-processing and Fairway.''  }

\vspace{-.25cm}

\begin{RQ}
{\bf RQ3.} How much labeled data is required to begin with?
\end{RQ}
Semi-supervised algorithms work with a small amount of labeled data and a large amount of unlabeled data. However, it is crucial to know how much labeled data is needed to start. Table \ref{Initial_size} shows results for three datasets where size of the \textit{initial fairly labeled set} has been varied from 1\% to 20\%. Here also we used the best version of Fair-SSL based on validation set results. We see the trend of increasing accuracy, F1, and decreasing AOD, EOD with increasing size of \textit{initial fairly labeled set}. Even when using 5\% labeled training data, we see Fair-SSL can significantly reduce bias. However, for accuracy and F1, Fair-SSL with 10\% labeled training data performs much better than 5\% version and very similar to 20\% version. Hence, to answer RQ3, we say that \textbf{``Fair-SSL can significantly reduce bias even when a very small amount of labeled data points are available. Performance of Fair-SSL gets better with increasing size of initial labeled training set.''}

\vspace{-.25cm}

\begin{RQ}
{\bf RQ4.} Which semi-supervised approach is the best to reduce bias? 
\end{RQ}
The results so far can be summarized as follows: 
semi-supervised algorithms can be useful to augment bias
mitigation process. That raises our next research question:
which one of our four different semi-supervised techniques performs
the best in this context?

Table \ref{RQ2_3} shows all four versions of Fair-SSL are reducing the bias scores. In case of performance metrics (the first five columns), all four of them are doing just the same (with minor differences). That means there is no way to choose one best method among the four methods. This reminds us of the popular ``No free lunch theorem'' for machine learning~\cite{nofreelunch}. We can certainly say semi-supervised pseudo-labeling improves fairness but which one is the best depends on the specific dataset.

We also compared the execution time of all four semi-supervised algorithms. We found out label propagation is the slowest and self-training is the fastest method. Hence, our answer for RQ4 is \textbf{``Semi-supervised techniques help but there is no winner based on performance. We should try all the approaches for a particular dataset to find out the best one. However, based on execution time, self-training works the fastest.''}

\section{Discussion \& Limitation}
\label{discussion}
What does make Fair-SSL unique and more useful than prior works?\\
\textbf{Inexpensive} - Real world data comes as a mixture of labeled and unlabeled form. Hiring Crowdsource workers for labeling can be very expensive~\cite{9064604}. Fair-SSL pseudo-labeler is a cheap alternative.\\
\textbf{Performance} - Fair-SSL can significantly reduce bias, improves recall \& F1 score and sometimes damages accuracy, false alarm. All the prior supervised works (except Fair-SMOTE~\cite{Chakraborty2021BiasIM}) damage performance to achieve fairness~\cite{Biswas_2020,kleinberg2016inherent,10.1145/3322640.3326705}. 
Performance wise Fair-SSL is similar to Fair-SMOTE but cheaper (uses 10\% data labels).\\ 
\textbf{Model-agnostic} - Fair-SSL is a data pre-processor. That means after pseudo-labeling the data, any supervised model can be used for final prediction. Thus Fair-SSL is model agnostic.

Fair-SSL does come with some limitations. We have used ten well-known datasets, three classification models, and four fairness metrics. Most of the prior works~\cite{NIPS2017_6988,Galhotra_2017,zhang2018mitigating,Kamiran:2018:ERO:3165328.3165686,chakraborty2019software} used one or two datasets and metrics. In future, we will explore more datasets and more learners. One assumption of evaluating our experiments is the test data is unbiased. Prior fairness studies also made similar assumption~\cite{Chakraborty2021BiasIM,Biswas_2020,chakraborty2019software}. SSL can be helpful when bias comes from improper data labels or sampling. But it can not solve some other bias causing factors such as objective function bias, homogenization bias~\cite{Das2020fairML}, and feature correlation bias~\cite{zhang2016causal}.

\section{Conclusion}
\label{conclusion}

Fairness in machine learning software has become a serious concern in the software engineering community. Prior fairness works mainly used supervised approaches that require data with ground truth labels. However, good quality labeled data is not always available in real-life. Keeping that in concern, this paper shows how semi-supervised techniques can be used to achieve fairness by using only 10\% labeled training data. We believe our work will encourage future software researchers to work in the software fairness domain.

% \section{Acknowledgement}
% \label{Acknowledgement}
% This research was funded by \textit{blinded for review}. 

\balance
\bibliographystyle{ACM-Reference-Format}
\bibliography{main}

\end{document}